\documentclass[%
reprint,
aps,
prb,
]{revtex4-2}
\usepackage{graphicx}
\usepackage{hyperref}
\usepackage{here}

\usepackage{color}
\usepackage{xcolor}
\usepackage{multirow}
\usepackage{dcolumn}
\usepackage{bm}
\usepackage{amsmath}
\usepackage{natbib}
\usepackage{docmute}
\usepackage{amssymb}
\usepackage{ulem}

\graphicspath{{./graph/}}

\usepackage{bm}

\newcommand{\imu}{{\rm i}}

\newcommand{\sgn}{\mathrm{sgn}\,}

\begin{document}

\title{
   Generalization of spectral bulk--boundary correspondence
}

\author{Shun Tamura$^{1}$, Shintaro Hoshino$^{2}$, and Yukio Tanaka$^{1}$}
\affiliation{%
$^1$Department of Applied Physics, Nagoya University, Nagoya 464-8603, Japan\\
$^2$Department of Physics, Saitama university, Saitama 338-8570, Japan
}

\date{\today}
\begin{abstract}
   The bulk--boundary correspondence in one dimension asserts that the physical quantities defined in the bulk and at the edge are connected, as well established in the argument for electric polarization.
   Recently, a spectral bulk--boundary correspondence (SBBC), an extended version of the conventional bulk--boundary correspondence to energy-dependent spectral functions, such as Green's functions, 
has been proposed in chiral symmetric systems, in which the chiral operator anticommutes with the Hamiltonian.
   In this study, we extend the SBBC to a system with impurity scattering and dynamical self-energies, regardless of the presence or absence of a gap in the energy spectrum.
Moreover, the SBBC is observed to hold even in a system without chiral symmetry, which substantially generalizes its concept.
The SBBC is demonstrated with concrete models, such as superconducting nanowires and a Su-Schrieffer-Heeger model.
Its potential applications and certain remaining issues are also discussed.
\end{abstract}
\pacs{pacs}

\maketitle
\section{Introduction}

Bulk--boundary correspondence (BBC) plays an 
important role in topological systems in which the physical quantities defined in the bulk and at the edge are connected.
This relation is useful because the physics at the surface can be determined from the bulk, which is theoretically easier to handle.
In topological materials, the number of zero-energy states appearing on the surface is predicted 
by an integer topological number, such as a Chern number or winding number defined in the bulk~\cite{PhysRevLett.49.405,KOHMOTO1985343,PhysRevLett.89.077002,RevModPhys.82.3045,PhysRevB.83.224511,RevModPhys.83.1057,RevModPhys.88.035005}.
The topological number is immune to a perturbation that does not break the symmetry of a Hamiltonian, and its value can be changed only through a topological quantum phase transition.
In addition, the BBC is known in the physics of electric polarization: the amount of accumulated charge 
is predicted by a geometric Berry phase 
defined in the bulk~\cite{PhysRevB.47.1651,PhysRevB.48.4442,RevModPhys.66.899}.
In contrast to the topological number, if there is no special symmetry, the electric polarization changes its value by perturbation, but the BBC still holds.
   For example, as discussed in Sec.~\ref{sec:SSH}, the Su-Schrieffer-Heeger model~\cite{PhysRevLett.42.1698,PhysRevB.22.2099} is characterized by the Berry phase defined in bulk which represents the polarization. The Berry phase is quantized as far as the chemical potential is zero due to the chiral symmetry.
   On the other hand, for the non-zero value of the chemical potential, the Berry phase is not quantized, but it still predicts the accumulated charge at the surface.

In chiral symmetric systems, where the Hamiltonian $H$ anticommutes with a chiral operator $\Gamma$, 
as $H\Gamma+\Gamma H=0$, a topological number called a winding number $W$ can be defined. It is related to the number of edge modes.
In our previous paper~\cite{PhysRevB.99.184512}, we reported that the BBC for chiral symmetric systems can be extended to a nonzero frequency $\omega$,
where the complex number of $W(\omega)$ is identical to a spectral function (i.e., Green's function) integrated over the edge region.
We refer to this relation as a spectral bulk--boundary correspondence (SBBC).
The SBBC was proved by Daido and Yanase~\cite{PhysRevB.100.174512}, by utilizing an analogy to the concept of electric polarization.
We also indicated that, for superconductors with chiral symmetry, an odd-frequency Cooper pair~\cite{doi:10.1143/jpsj.81.011013,RevModPhys.91.045005,Cayao2020}, which is characterized by a pair amplitude with an odd function in time or frequency, 
is induced at the surface and is related to the Green's functions defined in the bulk.
The SBBC has been demonstrated for mean-field systems, which are described as one-body Hamiltonians without retardation effects included in the self-energies.
Moreover, the impurity effect in the bulk has not been considered; however, it needs to be considered when comparing the results with those of real materials in which disorders are always present.

In this paper, we discuss the generalization of the SBBC\@.
We demonstrate that the SBBC holds for any amplitude of impurity in the bulk, after averaging the physical quantities over the impurity configurations.
The effect of the frequency-dependent self-energy that can appear due to tunneling~\cite{PMID:23665894} is also considered, where the SBBC still holds.
We further demonstrate that the SBBC can be applied to non-chiral-symmetric systems.
More specifically, although the existence of the chiral operator is necessary, 
it
does not necessarily anticommute with the Hamiltonian for the SBBC\@.
We also provide an outline of the proof for the generalized SBBC, along with the proof provided in Ref.~\cite{PhysRevB.100.174512}.

The remainder of this paper is organized as follows.
In Sec.~\ref{sec:sbbc}, we review the SBBC established in previous studies.
In Sec.~\ref{sec:generalize_sbbc}, we show how the SBBC is generalized and demonstrate it with several models.
We highlight the possible applications of the SBBC in Sec.~IV, and summarize our results in Sec.~\ref{sec:summary}.

\section{Spectral Bulk--Boundary Correspondence\label{sec:sbbc}}
This section briefly reviews the SBBC established in previous studies ~\cite{PhysRevB.99.184512,PhysRevB.100.174512}.
For topological systems, there is a relation 
between a topological number, which is defined in the bulk, and the number of zero-energy edge modes appearing at the edge of a semi-infinite system.
This nontrivial correspondence can be extended to a nonzero frequency for chiral symmetric systems.

For one-dimensional systems with chiral symmetry without many-body interaction, we can denote the Hamiltonian for a semi-infinite system as 
\begin{align}
   \hat{H}
   =&
   \sum_{i,j>0}
   \Psi^\dagger_{i} H_{i,j}\Psi_{j},
\label{eq:semiinf}
\end{align}
with $\Psi_{i}=(c_{i,1},\ldots,c_{i,m})$ where $c_{i,\alpha}$ 
($\alpha=1,\ldots,m$) is the annihilation operator for the $i$-th unit cell, and 
$\alpha$ indicates an index of the internal degree of freedom, such as a spin or orbital.
In this study, we set the lattice constant as a unit of length.
Equation~(\ref{eq:semiinf}) is used to calculate the physical quantities at the edge.
In contrast, the bulk quantities are calculated in the Hamiltonian with the periodic boundary condition given by
\begin{align}
   \hat{h}
   =
   \sum_k \psi_{k}^\dagger h(k)\psi_{k},
\end{align}
with $\psi_{k}=(c_{k,1},\ldots,c_{k,m})$ 
where $c_{k,\alpha}$ ($\alpha=1,\ldots,m$) is the annihilation operator 
for the momentum $k$.
Here, $H_{i,j}$ and $h(k)$ anticommute with $m\times m$
chiral operators $\Gamma$, where the relation
$\{ H_{i,j},\Gamma\}=\{h(k),\Gamma \}=0$ is satisfied with $\mathrm{Tr}\Gamma$=0~\cite{chiral}.
To avoid confusion, we use upper-case letters to denote the physical quantities in a semi-infinite system with a boundary at $j=0$, and 
lower-case letters to denote those in a periodic system designed for the bulk.

\begin{figure}[t]
   \centering
   \includegraphics[width=8.5cm]{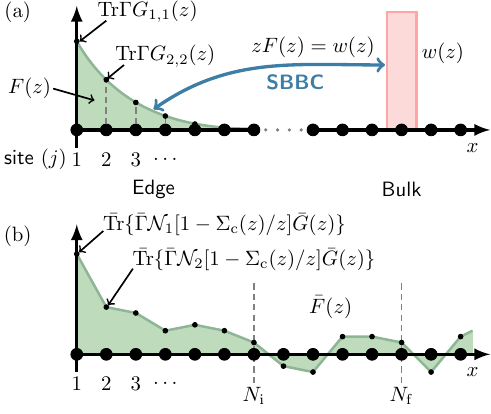}
   \caption{%
      Schematic illustrations of (a) SBBC given by Eq.~\eqref{eq:sbbc} [$F(z)$ is given by Eq.~\eqref{eq:f}] and
         (b) $\bar{F}(z)$ [$\bar{F}(z)$ is given by Eq.~\eqref{eq:def_bar_F} with Eq.~\eqref{eq:def_F_j}].
      The green area corresponds to $F(z)$ or $\bar{F}(z)$ and the red area shows $w(z)$.
   }%
   \label{fig:sbbc_pic}
\end{figure}

Then, the SBBC is explicitly written as
\begin{align}
   \label{eq:sbbc}
   zF(z)=w(z),
\end{align}
with 
\begin{align}
   \label{eq:f}
   F(z)
   &=
   \sum_{j>0}\mathrm{Tr}[\Gamma G_{j,j}(z)],\\
   \label{eq:w}
   w(z)
   &=
   \frac{1}{2\imu N_\mathrm{unit}}
   \sum_k
   \mathrm{Tr}[\Gamma g(k,z)\partial_k g^{-1}(k,z)],
\end{align}
where $N_\mathrm{unit}$ is the number of unit cells with sufficiently large values for convergence.
The trace $\mathrm{Tr}$ is taken in an $m\times m$ space.
Here, $G_{i,j}(z)$ and $g(k,z)$ are Green's functions given by $G_{i,j}(z)={(z-H)}_{i,j}^{-1}$ and 
$g(k,z)={[z-h(k)]}^{-1}$, respectively, where $z$ is a complex number that can be regarded as a generalized frequency.
Notably, $F(z)$ is an odd function of $z$ given by
\begin{align}
   F(z) =& \sum_{j>0}\mathrm{Tr}[\Gamma {(z-H)}^{-1}_{j,j}]
   \nonumber\\
   =&\sum_{j>0}\mathrm{Tr}[ {(z+H)}^{-1}_{j,j}\Gamma]
   \nonumber\\
   =&-F(-z).
   \label{eq:odd_f}
\end{align}
Then, $F(z)$ is given by the spatial sum of odd-frequency correlation functions in the edge region.
For superconductors, $F(z)$ can be the sum of anomalous Green's functions, and it is regarded as an odd-frequency Cooper pair amplitude~\cite{doi:10.1143/jpsj.81.011013,RevModPhys.91.045005,Cayao2020}.
A schematic of the SBBC is shown in Fig.~\ref{fig:sbbc_pic}(a).
Note that, in the limit $z\rightarrow0$ in Eq.~(\ref{eq:sbbc}), $zF(z)$ and $w(z)$ become
\begin{align}
   \label{eq:bec}
   \lim_{z\rightarrow0}zF(z)=\lim_{z\rightarrow0}w(z)=W,
\end{align}
where $W$ is an integer winding number~\cite{Gamma,PhysRevB.83.224511}, and 
Eq.~(\ref{eq:bec}) is a conventional bulk--boundary correspondence~\cite{physrevb.83.085426}.
Notably, $W$ can be well defined only when an energy gap at zero energy is opened.
From Eq.~(\ref{eq:bec}), the SBBC is regarded as an extension of the conventional bulk--boundary correspondence at zero frequency to a nonzero frequency.
   From Eqs.~\eqref{eq:sbbc}, \eqref{eq:odd_f} and \eqref{eq:bec} with sufficiently small value of $z$,
 $w(z)$ and $F(z)$ can be expanded as
   \begin{align}
      w(z)=&\sum_{l=0}^\infty {w}^{(l)}z^{2l},
      \label{eq:w_expansion}
      \\
      F(z)=&\sum_{l=0}^\infty {F}^{(l)} z^{2l-1},
      \label{eq:F_expansion}
   \end{align}
   as far as energy gap opens at zero energy.
   From the SBBC [Eq.~\eqref{eq:sbbc}], ${w}^{(l)}$ and ${\cal F}^{(l)}$ satisfy ${w}^{(l)}={F}^{(l)}$.
   It is notable that only ${w}^{(0)}$ and ${F}^{(0)}$ are related to the real topological number (${w}^{(0)}={F}^{(0)}=W$), and, ${w}^{(l)}$ ($={F}^{(l)}$) with $l>0$ is a complex number.
   Then, $w(z)$ and $F(z)$ are also a complex number.

Note that Eq.~(\ref{eq:w}) can be rewritten in real space by applying a Fourier transformation as follows:
\begin{align}
   w(z)
   =&
   \frac{1}{2\imu N_\mathrm{unit}}
   \mathrm{\bar{Tr}}\left\{\bar{\Gamma}g(z)\imu[g^{-1}(z),\bar{x}]\right\},
   \label{eq:w_real_space}
\end{align}
with
\begin{align}
   g_{i,j}(z)
   =&
   \frac{1}{N_\mathrm{unit}}
   \sum_k g(k,z) e^{\imu k(r_i-r_j)},
   \\
   h_{i,j}
   =&
   \frac{1}{N_\mathrm{unit}}
   \sum_k h(k) e^{\imu k(r_i-r_j)},
   \\
   \bar{\Gamma}_{i,j}
   =&
   \Gamma\delta_{i,j},
\end{align}
where $r_i$ is a spatial coordinate of the unit cell, and
the trace $\mathrm{\bar{Tr}}$ is taken for an $mN_\mathrm{unit}\times mN_\mathrm{unit}$ matrix.
Note that we have used the same symbol $g$ for both the real space and the $k$-space, and the symbols are distinguished by explicitly writing the arguments in parentheses.
 $\bar{x}$ is a \textit{position operator} defined by
\begin{align}
   \bar{x}
   =&
   \mathrm{diag}(r_1 I_{m},r_2 I_{m},\ldots,r_{N_\mathrm{unit}} I_{m}),
   \label{eq:position}
\end{align}
where $I_{m}$ is an $m\times m$ identity matrix~\cite{note_for_x,PhysRevLett.113.046802,PhysRevB.89.224203}.
As discussed in the following section, this real-space representation is crucial when we discuss the SBBC in the presence of impurities in the bulk, where the wavenumber is no longer a good quantum number unless we take the average over the impurity configuration.

It is notable that the surface Green's function $G_{1,1}(z)$ at small $|z|$ is connected to the topological number as discussed in Refs.~\cite{PhysRevB.85.165409,PhysRevB.95.235143}:
If there are zero-energy edge states, at least the diagonal elements of $G_{1,1}(z)$, which is related to the local density of states, diverges for small $z$, and if there is no zero-energy state, $G_{1,1}(z)$ is a regular function for small $z$.
Similarly in our case,
Eq.~\eqref{eq:bec} connects the winding number and the spatial sum of the Green's function $G_{j,j}(z)$ with $j>0$.
On the other hand, $w(z)$ can also be written only by the surface Green's functions which are defined in an open boundary condition system for both edges:
\begin{align}
   w(z)=
   \frac{1}{2}
   \mathrm{Tr}
   \Gamma
   \left[
      {(1-G_{N}tG_1t^{\dagger})}^{-1}
      -
      {(1-G_{1}t^\dagger G_{N}t)}^{-1}
   \right]
   \label{eq:w_surface}
\end{align}
with $G_{N}=G_{N_\mathrm{{unit}},{N_\mathrm{unit}}}(z)$, $G_{1}=G_{1,1}(z)$ and $t=H_{j,j+1}$.
Here the Green's function $G(z)$ is defined for the open boundary system with $N_\mathrm{unit}$ unit cells, and we 
have assumed $H_{j,j+l}=0$ with $l\geq2$ and translational symmetry in bulk (without impurity potential).
Eq.~\eqref{eq:w_surface} with $z\rightarrow0$ gives direct relation between the winding number and the surface Green's functions.
Details of the derivation is given in Appendix~\ref{sec:App_der_w}.

\section{Generalized SBBC\label{sec:generalize_sbbc}}
The SBBC can be applied to a system with impurities in the bulk and to a system without chiral symmetry, by changing its formula.
In Sec.~\ref{sec:formula_sbbc}, we provide a generalized version of the SBBC\@.
From Sec.~\ref{sec:kitaev_impurity} to~\ref{sec:SSH}, several examples of superconducting and nonsuperconducting systems in one dimension are studied.
In Sec.~\ref{sec:kitaev_impurity}, we discuss a Kitaev chain with impurities in which chiral symmetry is preserved.
In Sec.~\ref{sec:Rashba}, we discuss the self-energy effect on a Rashba nanowire proximately coupled to an $s$-wave superconductor.
In Sec.~\ref{sec:SSH}, a Su--Schrieffer--Heeger (SSH) model with chemical potential, which breaks the chiral symmetry, is discussed.

\subsection{Reformulation of SBBC\label{sec:formula_sbbc}}
This subsection considers a large but finite system with 
$N_\mathrm{unit}$ unit cells instead of a semi-infinite system.
We reformulate Eqs.~(\ref{eq:f}) and (\ref{eq:w}) for a system with impurities and self-energy.
First, we define the quantities at the boundary as follows:
\begin{align}
   \bar{F}(z)=&\langle F_j(z)\rangle_j,
\label{eq:def_bar_F}
\\
   F_j(z)=&
   \sum_{l=1}^{j}\bar{\mathrm{Tr}}
   \left\{{\cal N}_l\bar{\Gamma}\left[1-\Sigma_\mathrm{c}(z)/z\right] \bar{G}(z)\right\},
\label{eq:def_F_j}
   \\
   \bar{G}(z)=&{[z-H-\Sigma_\mathrm{c}(z)-\Sigma_\mathrm{ac}(z)]}^{-1},
   \label{eq:g_general}
\end{align}
where 
\begin{align}
   0=&
   \{\bar{\Gamma}, H\}=
   [\bar{\Gamma},\Sigma_\mathrm{c}(z)]
   =\{\bar{\Gamma},\Sigma_\mathrm{ac}(z)\},
   \label{eq:comm_anticomm1}
\end{align}
and $\langle\cdots\rangle_j=\frac{1}{N_\mathrm{f}-N_{\mathrm{i}}+1}\sum_{j=N_{\mathrm{i}}}^{N_{\mathrm{f}}}[\cdots]$, which performs self-averaging on the randomness of the impurity potentials~\cite{convergence}.
We choose $N_\mathrm{i}$ and $N_\mathrm{f}$ far from both edges [see Fig.~\ref{fig:sbbc_pic}(b) for a schematic], and we assume that $N_\mathrm{f}-N_\mathrm{i}$ is sufficiently large.
In addition, $\mathcal N_l$ is introduced by $({\cal N}_l)_{i,j} = I_m \delta_{i,l}\delta_{j,l}$.
Here, the suffixes ``c'' and ``ac'' indicate the commuting and anticommuting parts of the self-energy with the chiral operator, respectively.
Notably, any $2^n\times2^n$ (integer $n$) matrix can be decomposed by the products of the Pauli matrix, and any $2^n\times2^n$ matrix can be decomposed into commutable and anticommutable parts, as in Eq.~(\ref{eq:comm_anticomm1})~\cite{gell_mann}.

Here, the self-energy $\Sigma(z) \equiv \Sigma_\mathrm{c}(z)+\Sigma_\mathrm{ac}(z)$, which is divided into two parts, may be regarded as a part of the Hamiltonian if the $z$-dependence is absent.
In general, the presence of self-energy invalidates the chiral symmetry, which is defined as $\{ \bar{\Gamma}, H + \Sigma \} = 0$, or equivalently, $\Sigma_\mathrm{c}(z) =0$.
Nevertheless, the SBBC generally holds, as shown in the following section.

From the definition of $\bar{F}(z)$ in Eq.~\eqref{eq:def_bar_F}, the summation $\sum_{j=N_{\rm i}}^{N_{\rm f}}$ takes an average far from the edge and provides a mean value of $F_j(z)$ in the range
$F_{N_\mathrm{i}}(z)$ to $F_{N_\mathrm{f}}(z)$ [Fig.~\ref{fig:sbbc_pic}(b)].
If there is translational symmetry far from both edges, $F_j(z)$ is independent of $j$, as it is determined by the bulk properties.
In the limit $N_\mathrm{unit}\rightarrow\infty$ with $N_\mathrm{f}-N_\mathrm{i}\rightarrow\infty$
while keeping $N_\mathrm{i}$ and $N_\mathrm{f}$ sufficiently far from both edges, it can be expected that 
$z\bar{F}(z)$ converges to some value.

Next, we consider the quantities in the bulk.
The procedure is similar to the real-space representation discussed at the end of the previous section, but with the consideration of self-energy.
Namely, we define the quantities as
\begin{align}
   \bar{w}(z)
   =&
   \frac{1}{2\imu N_\mathrm{unit}}\bar{\mathrm{Tr}}\left\{\bar{\Gamma}\bar{g}(z)\imu [\bar{g}^{-1}(z),\bar{x}]\right\},
   \label{eq:bar_w}
   \\
   \bar{g}(z)=&
   {[z-h-\sigma_\mathrm{c}(z)-\sigma_\mathrm{ac}(z)]}^{-1},
\end{align}
with
\begin{align}
   0=&
   \{\bar{\Gamma}, h\}=
   [\bar{\Gamma},\sigma_\mathrm{c}(z)]
   =\{\bar{\Gamma},\sigma_\mathrm{ac}(z)\}.
   \label{eq:comm_anticomm2}
\end{align}
Note that the translational symmetry is not assumed in these expressions even in the bulk.

With the above definitions, the relation between $\bar{F}(z)$ and $\bar{w}(z)$ is given by
\begin{align}
   z\bar{F}(z)=&\bar{w}(z).
   \label{eq:sbbc_general}
\end{align}
which is one of the main results of this study.
We emphasize that Eq.~(\ref{eq:sbbc}) and Eq.~(\ref{eq:sbbc_general}) have the same functional form, but the latter can now be applied to non-chiral-symmetric and non-translation-symmetric systems.
See also Appendix~\ref{sec:App_proof_sbbc} for the derivation.
Notably, when $\Sigma_\mathrm{c}(z)\neq0$ or $\Sigma_\mathrm{ac}(z)\neq0$, the spectral function $\bar{F}(z)$ in the generalized SBBC can be a mixture of even and odd functions of $z$.
In the following subsections, we will demonstrate several relevant examples.

When there is an impurity potential,
we suppose that both the Hamiltonians with the open and periodic boundary conditions contain the impurity with the same impurity amplitude and probability distribution.
In Sec.~\ref{sec:kitaev_impurity} and Sec.~\ref{sec:SSH}, we use a uniform distribution of random numbers, and
adopt $N_\mathrm{i}=N_\mathrm{unit}/4$ and $N_\mathrm{f}=N_\mathrm{unit}/2$.

Without an impurity potential, $\bar{w}(z)$ can be written as
\begin{align}
   \bar{w}(z)=&
   \frac{1}{2\imu N_\mathrm{unit}}\sum_k \mathrm{Tr}[\Gamma \bar{g}(k,z)\partial_k \bar{g}^{-1}(k,z)],
\end{align}
where $\bar{g}(k,z)$ is obtained by applying Fourier transformation to $\bar{g}(z)$.
This recovers the original SBBC discussed in the previous section.
Notably, if $\sigma_{\mathrm{c}}(z)=0$, $\bar{w}(z=0)$ is an integer as long as the 
spectral gap remains open.

\subsection{Kitaev chain with impurity\label{sec:kitaev_impurity}}
We numerically demonstrate the impurity effect on the SBBC
for a Kitaev chain~\cite{kitaev_2001,PhysRevB.94.115166,PhysRevB.93.075129,universe5010033}, which is a model for a spin-polarized one-dimensional $p$-wave superconductor.
This model possesses chiral symmetry, and the Hamiltonian is given by
\begin{align}
   {\cal H}
   =&
   -t
   \sum_{j}
   \left(
      c_{j}^\dagger c_{j+1}
      +
      \mathrm{H.c.}
   \right)
   +
   \sum_j[\mu+tf_\mu(j)]
   c_j^\dagger c_j
   \nonumber\\
    &+
   \Delta 
   \sum_{j}
   \left(
      c_j^\dagger c_{j+1}^\dagger
      +
      \mathrm{H.c.}
   \right),
   \label{eq:h_kitaev_imp}
\end{align}
where $c_j$ ($c_j^\dagger$) is the annihilation (creation) operator for the $j$-th site,
$t$ is the hopping integral,
$\mu$ is the chemical potential, $\Delta$ is the superconducting pair potential, and
$f_\mu(j)$ is the impurity potential with $|f_\mu(j)|<V_\mu$.
The chiral operator is given by $\Gamma=\sigma_1$ with the Pauli matrix $\sigma_1$ in the particle--hole space.
To calculate $\bar{w}(z)$ and $\bar{F}(z)$, we use a recursive Green's function method~\cite{PhysRevB.55.5266} with the matrix transformation technique~\cite{eigen1977}, which enables us to access a large system.

\begin{figure}[t]
   \centering
   \includegraphics[width=8.5cm]{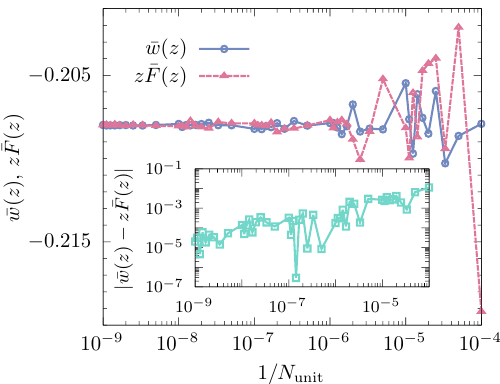}
   \caption{%
      Real part of $\bar{w}(z)$ and $z\bar{F}(z)$ plotted as a function of $1/N_\mathrm{unit}$
      for $\mu/t=0$, $\Delta/t=0.1$, $z/t=0.1\imu$, and $V_\mu=2$. 
      Imaginary part of $\bar{w}(z)$ and $z\bar{F}(z)$ are zero for purely imaginary $z$ due to chiral symmetry.
      $|\bar{w}(z)-z\bar{F}(z)|$ shown as a function of $1/N_\mathrm{unit}$ in the inset
   }%
   \label{fig:kitaev_imp_mu0}
\end{figure}

In Fig.~\ref{fig:kitaev_imp_mu0}, $\bar{w}(z)$ and $z\bar{F}(z)$ are shown as a function of the inverse of $N_\mathrm{unit}$ to examine the system size dependence.
Notably, we do not take a sample average, but prepare only one sample.
This graph shows that, with an increase in $N_\mathrm{unit}$, $\bar{w}(z)$ and $z\bar{F}(z)$ converge to the same value.
The difference between $z\bar{F}(z)$ and $\bar{w}(z)$ is also plotted in the inset of Fig.~\ref{fig:kitaev_imp_mu0}, 
and it decreases as $N_\mathrm{unit}$ increases.

Even in the presence of an impurity potential,
the winding number $W$ can be defined when an energy gap is opened~\cite{doi:10.1063/1.5026964}.
On the other hand, $\bar{w}(z)$ and $\bar{F}(z)$ are always well defined as long as $z$ is not equal to the eigenvalues of the Hamiltonian (which is avoided on the Matsubara axis), regardless of the presence or absence of an energy gap in the spectrum.
The impurity strength dependence of $\bar{w}(z)$ and $z\bar{F}(z)$ are discussed in the Appendix~\ref{sec:App_Kitaev}.

\begin{figure}[t]
   \centering
   \includegraphics[width=8cm]{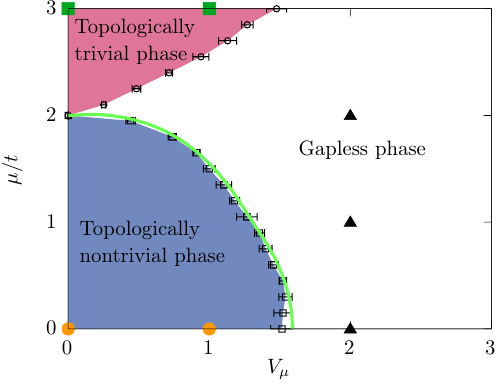}
   \caption{%
      Phase diagram for the Kitaev chain with impurity is shown as functions of $V_\mu$ and $\mu$ for $\Delta/t=0.1$.
      The phase boundary denoted by open squares and open circles  plots are determined by the eigen value of the Hamiltonian averaging over $10$ samples with $N_\mathrm{unit}=4000$ and the error bar indicates the standard deviation.
      The green line is determined by the transfer matrix method estimated by $10^8$ matrix product. 
      Red colored region is the topologically trivial phase, the blue colored region is the topologically nontrivial phase and other region is the gapless phase.
      Orange circles at $(V_\mu,\mu/t)=(0,0)$ and $(1,0)$ and green squares at $(0,3)$ and $(1,3)$ correspond to Fig.~\ref{fig:kitaev_imp_z_dep}(a) and (b), respectively.
      Black triangles at $(V_\mu,\mu/t)=(2,0)$, $(2,1)$ and $(2,2)$ correspond to Fig.~\ref{fig:kitaev_imp_z_dep}(c) and (d).
   }%
   \label{fig:phase_kitaev}
\end{figure}
We show the phase diagram for the Hamiltonian Eq.~\eqref{eq:h_kitaev_imp} in Fig.~\ref{fig:phase_kitaev} [see also Appendix~\ref{sec:App_Kitaev_dos}].
There are three phases: topologically trivial, nontrivial and gapless phases.
We also calculate an inverse participation ratio and all the wave functions are localized for $V_\mu>0$ even in the gapless phase [Appendix~\ref{sec:App_Kitaev_IPR}].
This result is consistent with the Anderson localization.
In principle, the phase boundary can be determined by the gap closing points
estimated by the eigenvalues of the Hamiltonian.
We evaluate the energy gap by calculating the eigen values of Eq.~\eqref{eq:h_kitaev_imp} 
and show the phase boundary by open squares and open circles in Fig.~\ref{fig:phase_kitaev}.
The phase boundary between the topologically nontrivial and the gapless phase can also be obtained
by the transfer matrix method~\cite{PhysRevB.93.075129} (shown by green curve) and this boundary is almost the same as estimated by gap closing points.

\begin{figure}[t]
   \centering
   \includegraphics[width=8.5cm]{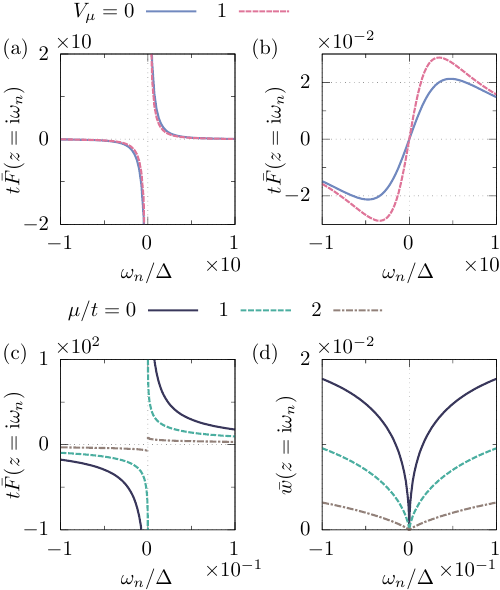}
   \caption{%
      (a) and (b) imaginary part of $\bar{F}(z=i\omega_n)$ is plotted as a function of $\omega_n$ for $V_\mu=0$ and $V_\mu=1$.
      (a) $\mu=0$ (indicated by orange circles in topologically nontrivial phase in 
      Fig.~\ref{fig:phase_kitaev}) (b) $\mu/t=3$ (indicated by green squares in topologically trivial phase in 
      Fig.~\ref{fig:phase_kitaev})
      In (c) and (d), imaginary part of $\bar{F}(z=\mathrm{i}\omega_n)$ and real part of $\bar{w}(z=\mathrm{i}\omega_n)$, respectively are shown for $\mu/t=0$ to $2$ with $V_\mu=2$ (indicated by black triangles in the gapless phase in Fig.~\ref{fig:phase_kitaev}).
      In all the plots, we choose $\Delta/t=0.1$ and $N_\mathrm{unit}=10^8$.
   }%
   \label{fig:kitaev_imp_z_dep}
\end{figure}
In Figs.~\ref{fig:kitaev_imp_z_dep}(a)--(c), $\omega_n$ dependence of $\bar{F}(\mathrm{i}\omega_n)$ is shown. 
As can be seen in the graphs, $\bar{F}(\mathrm{i}\omega_n)$ is an odd function of $\omega_n$ since the impurity potential does not break chiral symmetry.
Then $\bar{F}(\mathrm{i}\omega_n)$ can be regarded as an odd-frequency Cooper pair amplitude~\cite{doi:10.1143/jpsj.81.011013,RevModPhys.91.045005,Cayao2020}
even in the presence of the impurity potential.
In Fig.~\ref{fig:kitaev_imp_z_dep}(a), $\bar{F}(\mathrm{i}\omega_n)$ for topologically nontrivial phase is shown. 
It is notable that 
$\bar{w}(z)$ and $\bar{F}(z)$ can be expanded like Eq.~\eqref{eq:F_expansion} in the gapped phases (topologically nontrivial and trivial phase) due to chiral symmetry:
\begin{align}
   \bar{w}(z)=&\sum_{l=0}^\infty \bar{w}^{(l)} z^{2l},
   \label{eq:bar_w_expansion}
   \\
   \bar{F}(z)=&\sum_{l=0}^\infty \bar{F}^{(l)} z^{2l-1},
   \label{eq:bar_F_expansion}
\end{align}
with $\bar{w}^{(l)}=\bar{F}^{(l)}$.
Here, $|z|$ is smaller than the energy gap.
For the topologically nontrivial phase, energy gap opens and $\bar{w}(z=0)=\lim_{z\rightarrow0}z\bar{F}(z)=W$ i.e., $\bar{w}^{(0)}=\bar{F}^{(0)}=W$ holds even in the presence of the impurity potential due to chiral symmetry.
Therefore, $\bar{F}(\mathrm{i}\omega_n)$ diverges in the topologically nontrivial phase due to nonzero value of $W$.
Difference between $\bar{F}(\mathrm{i}\omega_n)$ for $V_\mu=0$ and that for $1$ is not very large [Fig.~\ref{fig:kitaev_imp_z_dep}(a)].

In Fig.~\ref{fig:kitaev_imp_z_dep}(b), $\bar{F}(\mathrm{i}\omega_n)$ is shown for the topologically trivial phase. 
For small value of $\omega_n$, $\bar{F}(\mathrm{i}\omega_n)$ is a linear function of $\omega_n$
since $\bar{F}^{(0)}=W=0$ in Eq.~\eqref{eq:bar_F_expansion} for the topologically trivial phase.
The slope close to $\omega_n=0$ is given by $\bar{F}^{(1)}$ and $\bar{F}^{(1)}$ becomes larger as $V_\mu$ becomes larger in this phase.

In the gapless phase, we cannot apply Laurent expansion like Eqs.~\eqref{eq:bar_w_expansion} and \eqref{eq:bar_F_expansion} since the radius of convergence is zero at $z=0$ for $N_\mathrm{unit}\rightarrow\infty$ where nearly-zero eigenvalues of the Hamiltonian are present.
In the numerical calculation, for smaller value of $z$, we have to take larger system size $N_\mathrm{unit}$.
We show $\bar{F}(\mathrm{i}\omega_n)$ in Fig.~\ref{fig:kitaev_imp_z_dep}(c).
Within our numerical calculation, for $\mu/t=0$ and $\mu/t=1$, $\bar{F}(\mathrm{i}\omega_n)$ diverges for small $\omega_n$ and at $\mu/t=2$, $\bar{F}(\mathrm{i}\omega_n)$ approaches a finite value for $\omega_n\rightarrow0$ (We numerically confirm this up to $\omega_n/\Delta=10^{-5}$).
The low-frequency behaviors are more clearly 
seen from $\bar{w}(\mathrm{i}\omega_n)$ in Fig.~\ref{fig:kitaev_imp_z_dep}(d).
In $\omega_n\rightarrow0$ limit, $\bar{w}(\mathrm{i}\omega_n)$ approaches zero and from Fig.~\ref{fig:kitaev_imp_z_dep}(d), which implies that $\bar{w}(\mathrm{i}\omega_n)$ cannot be expanded like Eq.~\eqref{eq:bar_w_expansion}:
$\bar{w}(\mathrm{i}\omega_n)$ behaves 
as
$\bar{w}(\mathrm{i}\omega_n)\sim{|\omega_n|}^\kappa$ with $0<\kappa\lesssim1$ for small $\omega_n$.
Then, from the SBBC, $\bar{F}(\mathrm{i}\omega_n)$ can be written as $\bar{F}(\mathrm{i}\omega_n)=\bar{w}(\mathrm{i}\omega_n)/(\mathrm{i}\omega_n)\sim-\mathrm{i}{|\omega_n|}^{\kappa-1}\sgn(\omega_n)$.
Within our numerical calculation, for $\mu/t=0$ and $1$ in Fig.~\ref{fig:kitaev_imp_z_dep}(d), $0<\kappa<1$ is satisfied and then $\bar{F}(\mathrm{i}\omega_n)$ diverges for $\omega_n\rightarrow0$.
For $\mu/t=2$, we obtain $\kappa\sim1$  and then $\bar{F}(\mathrm{i}\omega_n)$ approaches a finite value for $\omega_n\rightarrow0$.

\subsection{Rashba nanowire\label{sec:Rashba}}
\begin{figure}[t]
   \centering
   \includegraphics[width=8.5cm]{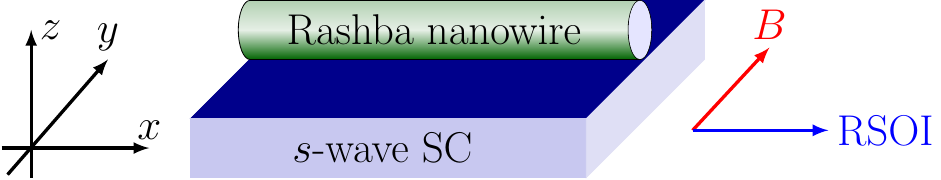}
   \caption{%
      Schematic of normal metal Rashba nanowire on $s$-wave superconductor (SC); 
      RSOI represents Rashba spin--orbit interaction.
   }%
   \label{fig:nanowire_pic}
\end{figure}

In this subsection, we consider a Green's function with self-energy for a Rashba nanowire system~\cite{PhysRevLett.105.077001,PhysRevLett.105.177002,Mourik1003}.
The Rashba nanowire system has attracted considerable attention because it hosts Majorana fermions at both edges.
Figure~\ref{fig:nanowire_pic} illustrates a Rashba nanowire system that possesses 
a large Rashba spin--orbit interaction and is proximately coupled to an $s$-wave superconductor.
The Hamiltonian for the Rashba nanowire system is 
\begin{align}
   {\cal H}_{\mathrm{RN}}
   =&
   -t\sum_{j,\sigma}
   \left(
      c_{j,\sigma}^{\dagger}
      c_{j+1,\sigma}
      +\mathrm{H.c.}
   \right)
   -\mu\sum_{j,\sigma}c_{j,\sigma}^{\dagger}c_{j,\sigma}
   \nonumber\\
    &+B\sum_{j,\sigma,\sigma'}
      c_{j,\sigma}^{\dagger}
      {(\hat{\sigma}_{3})}_{\sigma,\sigma'}
      c_{j,\sigma'}
   \nonumber\\
    &+\frac{i\alpha}{2}
   \sum_{j,\sigma,\sigma'}
   \left[
      c_{j,\sigma}^{\dagger}
      {(\hat{\sigma}_{2})}_{\sigma,\sigma'}
      c_{j+1,\sigma'}
      -\mathrm{H.c.}
   \right],
   \label{eq:hamiltonian_rn}
\end{align}
with a Pauli matrix $\hat{\sigma}_i$ ($i=1,2,3$) in the spin space.
Here, $B$ is the magnetic field, and $\alpha$ is the Rashba spin--orbit interaction.
The $s$-wave pair potential is included in the self-energy~\cite{PMID:23665894} and is given by
\begin{align}
   \Sigma(\mathrm{i}\omega_n)=&
   -|\tilde{t}|^2\nu_\mathrm{F}
   \left[
      \frac{\omega_n+\Delta_s\hat{\sigma}_2\hat{\tau}_2}{\sqrt{\Delta_s^2-\omega_n^2}}+\zeta\hat{\tau}_3
   \right],
   \label{eq:self_energy}
\end{align}
where $\tilde{t}$ is the hopping between the nanowire and the superconductor, $\nu_\mathrm{F}$ is the density of states (DOS) at the surface of the superconductor at the Fermi energy,
$\omega_n$ is the Matsubara frequency,
$\Delta_s$ is the $s$-wave pair potential of the superconductor, $\zeta$ is the proximity-induced chemical potential shift,
and $\hat{\tau}_i$ ($i=1,2,3$) is a Pauli matrix in the particle--hole space.
Namely, the retardation effect associated with electron tunneling between the nanowire and the superconductor is expressed as the frequency-dependent self-energy.

The chiral operator in Eq.~(\ref{eq:hamiltonian_rn}) is $\Gamma=\hat{\tau}_1$, and the self-energy in 
Eq.~(\ref{eq:self_energy}) can be decomposed as $\Sigma(\imu\omega_n)=\Sigma_\mathrm{c}(\imu\omega_n)+\Sigma_\mathrm{ac}(\imu\omega_n)$ with
\begin{align}
   \Sigma_\mathrm{c}(\imu\omega_n)
   =&
   -|\tilde{t}|^2\nu_\mathrm{F}
   \frac{\omega_n}{\sqrt{\Delta_s^2-\omega_n^2}},
   \label{eq:self_nano_c}
   \\
   \Sigma_\mathrm{ac}(\imu\omega_n)
   =&
   -|\tilde{t}|^2\nu_\mathrm{F}
   \left[
      \frac{\Delta_s\hat{\sigma}_2\hat{\tau}_2}{\sqrt{\Delta_s^2-\omega_n^2}}+\zeta\hat{\tau_3}
   \right].
   \label{eq:self_nano_ac}
\end{align}
For the open boundary system, self-energy can have site dependence close to the surface, although it is independent of the site for the bulk.
As discussed in Refs.~\cite{PhysRevB.99.184512,PhysRevB.100.174512}, $\bar{F}(z)$ does not change its value 
with the surface modulation of the Hamiltonian or the self-energy.
Therefore, we can use the value of the self-energy in the bulk, even at the boundary.
From the above Hamiltonian and self-energy, we can explicitly write the SBBC relation~\eqref{eq:sbbc_general}.

For the Rashba nanowire system, the Green's function $\bar{g}(z)$ satisfies $\bar{\Gamma} \bar{g}(z) \bar{\Gamma}=-\bar{g}(-z)$ with $\bar{\Gamma}=\mathrm{diag}(\hat{\sigma}_0\hat{\tau}_1,\hat{\sigma}_0\hat{\tau}_1,\ldots)$ and then, the winding number is given by $W=\bar{w}(z=0)$~\cite{PhysRevB.86.165116,PhysRevB.86.205119}.
Here we use the relations $\Sigma_{\mathrm{c}}(-z)=-\Sigma_{\mathrm{c}}(z)$ and $\Sigma_{\mathrm{ac}}(-z)=\Sigma_{\mathrm{ac}}(z)$ from Eqs.~\eqref{eq:self_nano_c} and \eqref{eq:self_nano_ac}.
Then the phase boundary is given by 
$B_1=\sqrt{{(\mu+2t+|\tilde{t}|^2\nu_\mathrm{F}\zeta)}^2+{(|\tilde{t}|^2\nu_\mathrm{F})}^2}$ and
$B_2=\sqrt{{(-\mu+2t+|\tilde{t}|^2\nu_\mathrm{F}\zeta)}^2+{(|\tilde{t}|^2\nu_\mathrm{F})}^2}$.
The topologically nontrivial phase is 
$B_\mathrm{min}<|B|<B_\mathrm{max}$ with $B_\mathrm{min}=\min(B_1,B_2)$ and $B_\mathrm{max}=\max(B_1,B_2)$.

\begin{figure}[t]
   \centering
   \includegraphics[width=8.5cm]{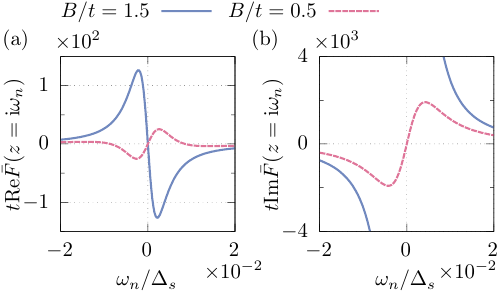}
   \caption{%
      (a) Real part of $\bar{F}(z=\mathrm{i}\omega_n)$ and (b) imaginary part of it is plotted as a function of $\omega_n$ for the topologically nontrivial ($B/t=1.5$) and trivial phase ($B/t=0.5$) with $\mu/t=-1$, $\alpha/t=0.5$, $\Delta_s/t=0.01$, $\tilde{t}/t=0.01$, $\nu_\mathrm{F}t=1$ and $\zeta/t=0.1$.
   }%
   \label{fig:nano_wire_F}
\end{figure}

As mentioned in Sec.~\ref{sec:formula_sbbc}, in general, $\bar{F}(z)$ is the combination of even- and odd-frequency components.
However, in this case, $\bar{F}(z)$ is purely odd-frequency due to the specific structure of the self-energy [Eqs.~\eqref{eq:self_nano_c} and \eqref{eq:self_nano_ac}]:
\begin{align}
   F_j(-z)
   =&
   \sum_{l=1}^j\bar{\mathrm{Tr}}
   \left\{{\cal N}_l\bar{\Gamma}[1+\Sigma_{\mathrm{c}}(-z)/z]\bar{G}(-z)\right\}
   \nonumber\\
   =&
   \sum_{l=1}^j\bar{\mathrm{Tr}}
   \Big\{{\cal N}_l[1-\Sigma_{\mathrm{c}}(z)/z]
      \nonumber\\
    &
    \times\bar{\Gamma}
 {[-z-H-\Sigma_\mathrm{c}(-z)-\Sigma_\mathrm{ac}(-z)]}^{-1}\Big\}
 \nonumber\\
   =&
   -
   \sum_{l=1}^j\bar{\mathrm{Tr}}
   \left\{{\cal N}_l[1-\Sigma_{\mathrm{c}}(z)/z]\bar{G}(z)\bar{\Gamma}\right\}
   \nonumber\\
   =&
   -F_j(z).
   \label{eq:F_nanowire}
\end{align}
This can also be seen from Figs.~\ref{fig:nano_wire_F}(a) and (b).
It is noted that for the topologically nontrivial phase, the imaginary part of $\bar{F}(z=\mathrm{i}\omega_n)$ diverges at $\omega_n=0$ due to the presence of the zero-energy state.
It is noted that from Eq.~\eqref{eq:F_nanowire}, $\bar{F}(z)$ can be expanded like Eq.~\eqref{eq:bar_F_expansion}. Then, the real part of $\bar{F}(\mathrm{i}\omega_n)$ is a linear function at small $\omega_n$ [Fig.~\ref{fig:nano_wire_F}(a)] since $\bar{F}^{(0)}=W$ is an integer and from Eq.~\eqref{eq:bar_F_expansion}, $\mathrm{Re}\bar{F}(\mathrm{i}\omega_n)=\mathrm{Re}\sum_{l=1}^\infty\bar{F}^{(l)}{(\mathrm{i}\omega_n)}^{2l-1}$ that the $l=0$ term vanishes and the summation starts from $l=1$.
In the topologically trivial phase, $\bar{F}(\mathrm{i}\omega_n)$ is a linear function of $\omega_n$ due to the absence of the zero-energy state~\cite{PhysRevB.99.184512}.
   It is also noted that in general, without the self-energy, $F(\mathrm{i}\omega_n)$ is a purely imaginary function.
   On the other hand, in the presence of the self-energy, $\bar{F}(\mathrm{i}\omega_n)$ has non-zero value of both real and imaginary components.

When the direction of the magnetic field is perpendicular to the Rashba spin--orbit interaction, 
the Hamiltonian Eq.~(\ref{eq:hamiltonian_rn}) has chiral symmetry, 
and there are Majorana fermions at both edges in a topological phase.
Notably, when their directions are not perpendicular to each other, 
these Majorana fermions are still present up to a certain angle~\cite{PhysRevB.90.115429},
and the SBBC also holds.

\subsection{DOS and SBBC for SSH model\label{sec:SSH}}
\subsubsection{Without impurity}
The concept of SBBC can also be applied to nonsuperconducting systems.
In this subsection, we first discuss the relationship between $F(z)$ and the DOS for an SSH model with chiral symmetry,
which is a model for polyacetylene~\cite{PhysRevLett.42.1698,PhysRevB.22.2099}. 
Then, we discuss the SBBC for an SSH model with a nonzero value of chemical potential, which breaks the chiral symmetry.

The SSH model has two sublattices in the unit cell.
The Hamiltonian is given by
\begin{align}
   {\cal H}_{\mathrm{SSH}}
   =&
   -t_1\sum_j
   \left(
      c_{j,\mathrm{A}}^\dagger
      c_{j,\mathrm{B}}
      +
      \mathrm{H.c.}
   \right)
   \nonumber\\
     &
   -t_2\sum_j
   \left(
      c_{j,\mathrm{B}}^\dagger
      c_{j+1,\mathrm{A}}
      +
      \mathrm{H.c.}
   \right)
   \label{eq:hamiltonian_ssh1}
   \\
   =&
   \sum_k
   \begin{pmatrix}
      c_{k,\mathrm{A}}^\dagger&
      c_{k,\mathrm{B}}^\dagger
   \end{pmatrix}
   h(k)
   \begin{pmatrix}
      c_{k,\mathrm{A}}\\
      c_{k,\mathrm{B}}
   \end{pmatrix},
   \label{eq:hamiltonian_ssh2}
   \\
   h(k)=&
   \left[
      -(t_1+t_2\cos k)\tau_1
      -t_2\sin k\tau_2
   \right],
   \label{eq:hamiltonian_ssh3}
\end{align}
where A and B denote the sublattices, and $\tau_j$ with $j=1,2,3$ is a Pauli matrix in the sublattice space.
Here, the chiral operator is $\Gamma=\tau_3$, and only the diagonal elements of the chiral operator are nonzero.
The SSH model becomes topologically nontrivial when $t_2/t_1>1$ for non-negative $t_1$ and $t_2$, as characterized by the nonzero winding number.
The phase boundary is characterized by the winding number $W$ and the Berry phase $\theta$:
\begin{align}
   \theta=i\sum_{n\in\mathrm{occ}}\int_{-\pi}^\pi dk \langle u_{n,k}|\partial_k u_{n,k}\rangle.
   \label{eq:berry_phase}
\end{align}
Here, summation is taken for occupied states and $|u_{n,k}\rangle$ is an eigen state of the Hamiltonian $h(k)$.
$W$ and $\theta$ satisfy $|W|=|1-e^{i\theta}|/2$, i.e., $\theta=(2n+1)\pi$ for $t_2/t_1>1$ and $\theta=2n\pi$ for $t_2/t_1<1$ ($n\in\mathbb{Z}$).
For the SSH model, the Berry phase is related to the polarization~\cite{PhysRevB.47.1651,PhysRevB.48.4442,RevModPhys.66.899}. 

From the SBBC with $z=\imu\omega_n$, $F(\imu\omega_n)$ is related to 
\textit{an odd-frequency charge density wave} as follows~\cite{PhysRevB.101.214507}:
\begin{align}
   F(\imu\omega_n)=\int_0^{\beta}d\tau e^{\imu\omega_n\tau}\sum_{j>0}
   \langle
   c_{j,\mathrm{A}}^\dagger(\tau)
   c_{j,\mathrm{A}}(0)
   -
   c_{j,\mathrm{B}}^\dagger(\tau)
   c_{j,\mathrm{B}}(0)
   \rangle.
\end{align}

Through the analytic continuation,
$F(z=E+\imu\eta)$ with a positive infinitesimal $\eta$ is related to the difference in the DOS between the A and B sublattices 
because the local DOS is given by the imaginary part of the retarded Green's function, $\rho(E,j)=-\frac{1}{\pi}\mathrm{Im}G_{j,j}(z=E+\imu\eta)$:
\begin{align}
   -\frac{1}{\pi}\mathrm{Im}F(E+\imu\eta)
   =&
   \rho_\mathrm{A}(E)
   -
   \rho_\mathrm{B}(E).
   \label{eq:ssh_dos}
\end{align}
Here, $\rho_{\mathrm{A}(\mathrm{B})}$ is the DOS at the A(B) sublattice given by
$\rho_{\mathrm{A}(\mathrm{B})}(E)=-\frac{1}{\pi}\mathrm{Im}\sum_{j=1}^{N_\mathrm{f}}G_{(j,\mathrm{A}(\mathrm{B})),(j,\mathrm{A}(\mathrm{B}))}(E+\imu\eta)$.
Equation~(\ref{eq:ssh_dos}) can be derived from $\Gamma=\tau_3$ as follows:
\begin{align}
   F(z)
   =
   \sum_{j>0} \left[G_{(j,\mathrm{A}),(j,\mathrm{A})}(z)-G_{(j,\mathrm{B}),(j,\mathrm{B})}(z)\right].
    \label{eq:F_ssh}
\end{align}
In general, $\rho_\mathrm{A}(E)$ and $\rho_\mathrm{B}(E)$ diverge since the local density of states are not zero in bulk and then the spatial sum does not converge for $N_{\mathrm{f}}\rightarrow \infty$ i.e., $\rho_\mathrm{A(B)}(E)\propto N_{\mathrm{f}}$.
However, $\rho_\mathrm{A}(E)-\rho_\mathrm{B}(E)$ (except for $E=0$, $E=\pm|t_1\pm t_2|$ for the topologically nontrivial phase and $E=\pm|t_1\pm t_2|$ for the topologically trivial phase) converges due to the SBBC\@.
Then, from the SBBC, the difference in the DOS between the A and B sublattices for the semi-infinite system
can be predicted from the bulk value $w(E+\imu\eta)$ (see Appendix~\ref{sec:App_SSH}).
Figure~\ref{fig:SSH}(a) shows $\rho_\mathrm{A}(E)-\rho_\mathrm{B}(E)$.
In the topologically nontrivial phase, $\rho_\mathrm{A}(E)-\rho_\mathrm{B}(E)$ 
diverges at $E=0$ [Fig.~\ref{fig:SSH}(a) with $t_2/t_1=1.5$].
   In the inset of Fig.~\ref{fig:SSH}(a), we show the magnified view close to $E=0$. 
   Here, the peak width depends on the value of $\eta$.
\begin{figure}[H]
   \centering
   \includegraphics[width=8.5cm]{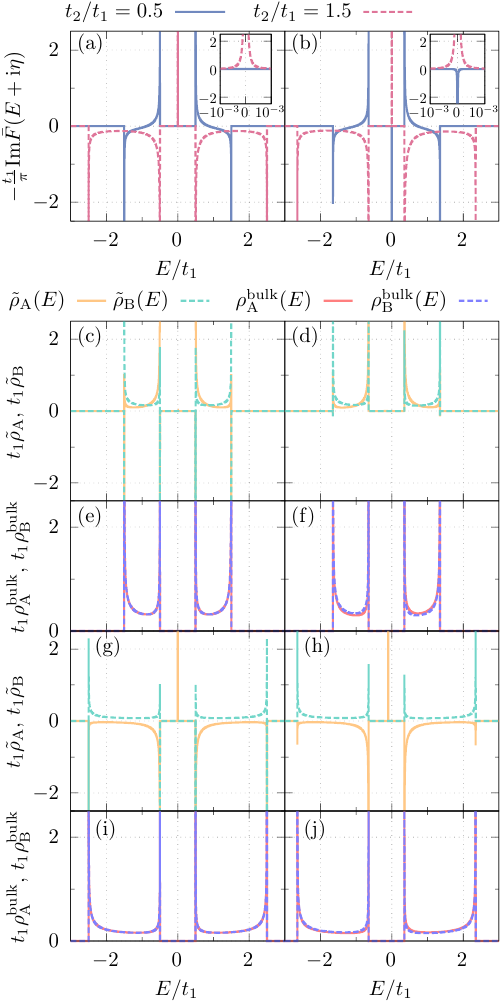}
   \caption{%
      Difference in the local DOS plotted as a function of 
      $E$ for $t_2/t_1=0.5$ and $1.5$ (a) $\mu_\mathrm{A}=\mu_\mathrm{B}=0$ [Eq.~\eqref{eq:ssh_dos}] (b) $\mu_\mathrm{A}/t_1=0.1$ and $\mu_\mathrm{B}/t_1=0.2$ [Eq.~\eqref{eq:F_E_ssh_general}]. 
         Magnified views close to $E=0$ are shown in the insets.
         $\tilde{\rho}_\mathrm{A(B)}(E)$ [Eq.~\eqref{eq:tilde_rho_ssh}] is plotted as a function of $E$ in (c), (d), (g) and (h) and
         ${\rho}_\mathrm{A(B)}^\mathrm{bulk}(E)$ is plotted as a function of $E$ in (e), (f), (i) and (j).
         (c) and (e) $(t_2/t_1,\mu_\mathrm{A}/t_1,\mu_\mathrm{B}/t_1)=(0.5,0,0)$ (d) and (f) $(0.5,0.1,0.2)$ (g) and (i) $(1.5,0,0)$ (h) and (j) $(1.5,0.1,0.2)$.
         In all plots, $\eta/t_1=10^{-7}$
   }%
   \label{fig:SSH}
\end{figure}
For superconductors, $F(z)$ is related to an anomalous Green's function and is difficult to detect because
the anomalous Green's function is a gauge-dependent quantity.
Conversely, $F(z)$ for the SSH model is composed of a normal Green's function, which can be measured.

In Figs.~\ref{fig:SSH}(c) and (g), we show the difference between $\rho_\mathrm{A(B)}(E)$ and the DOS in bulk:
\begin{align}
      &\tilde{\rho}_\mathrm{A(B)}(E)
      =
      \frac{-1}{\pi}
      \mathrm{Im}
      \phi_{\mathrm{A(B)}}(E+\mathrm{i}\eta),
      \label{eq:tilde_rho_ssh}
      \\
      &\phi_{\mathrm{A(B)}}(z)
      =
      \sum_{j=1}^{N_\mathrm{f}}
      \left[
         G_{(j,\mathrm{A(B)}),(j,\mathrm{A(B)})}(z)-g_{\mathrm{A(B)},\mathrm{A(B)}}(z)
      \right].
\end{align}
Here, $g_{\mathrm{A(B)},\mathrm{A(B)}}(z)$ is a local Green's function in bulk at A(B) sublattice.
In general, $\rho_\mathrm{A(B)}(E)$ diverges since the DOS in bulk is not zero but $\tilde{\rho}_\mathrm{A(B)}(E)$ is finite except for van Hove singularity points.
When the chemical potential is zero, the DOS in bulk satisfy $\rho_\mathrm{A}^\mathrm{bulk}(E)=\rho_\mathrm{B}^\mathrm{bulk}(E)$ due to chiral symmetry [Figs.~\ref{fig:SSH}(e) and (i)]. Here $\rho_\mathrm{A(B)}^\mathrm{bulk}(E)=-\frac{1}{\pi}\mathrm{Im}g_{\mathrm{A(B),A(B)}}(E+\mathrm{i}\eta)$ is the DOS in bulk at A(B) sublattice.
Therefore, Eq.~\eqref{eq:ssh_dos} can also be written as $-\frac{1}{\pi}\mathrm{Im}F(E+\mathrm{i}\eta)=\tilde{\rho}_\mathrm{A}(E)-\tilde{\rho}_\mathrm{B}(E)$ for zero chemical potential.
Then, Fig.~\ref{fig:SSH}(a) can be understood from Figs.~\ref{fig:SSH}(c) and (g).

In Eqs.~(\ref{eq:hamiltonian_ssh1}), (\ref{eq:hamiltonian_ssh2}), and (\ref{eq:hamiltonian_ssh3}),
the chemical potentials are set to zero, and the Hamiltonian has chiral symmetry, that is, it anticommutes with the chiral operator.
On the other hand, in a more realistic situation, the chemical potentials are not zero. 
\begin{align}
   {\cal H}=&{\cal H}_\mathrm{SSH}
   -\mu_\mathrm{A}\sum_j n_{j,\mathrm{A}}
   -\mu_\mathrm{B}\sum_j n_{j,\mathrm{B}},
\end{align}
with $n_{j,\mathrm{A(B)}}=c_{j,\mathrm{A(B)}}^\dagger c_{j,\mathrm{A(B)}}$~\cite{PhysRevB.89.085111}.
For the non-zero value of $\mu_\mathrm{A}$ or $\mu_\mathrm{B}$, there is not chiral symmetry.
In this case, the winding number cannot be defined and the Berry phase given by Eq.~\eqref{eq:berry_phase} is 
not quantized: $\theta\neq n\pi$.
However, the SBBC still holds, which is similar to the fact that the Berry phase predicts a surface charge~\cite{PhysRevB.47.1651,PhysRevB.48.4442,RevModPhys.66.899}.
We can choose any $\tau_{i=1,2,3}$ for $\Gamma$, indicating that there are several SBBC relations for a given physical system.
More generally, we can choose the superposition of these Pauli matrices in principle (see Sec.~\ref{sec:discussion}).
When we choose $\Gamma=\tau_3$, 
\begin{align}
   \Sigma_{\mathrm{c}}(z)
   =&
   -\mu_\mathrm{A}\sum_j n_{j,\mathrm{A}}
   -\mu_\mathrm{B}\sum_j n_{j,\mathrm{B}},
   \\
   \sigma_{\mathrm{c}}(z)=&
   -\mu_\mathrm{A}(\tau_0+\tau_3)/2 - \mu_\mathrm{B}(\tau_0-\tau_3)/2.
\end{align}
Notably, the extended version of the winding number $\bar{w}(z)$ with $z=0$ is not an integer for the nonzero 
value of $\mu_\mathrm{A}$ or $\mu_\mathrm{B}$.
Then, $\bar{F}(z)$ is
\begin{align}
   \bar{F}(z)
   =&
   \sum_{j=1}^{N_\mathrm{f}} \left[\bar{G}_{(j,\mathrm{A}),(j,\mathrm{A})}(z)-\bar{G}_{(j,\mathrm{B}),(j,\mathrm{B})}(z)\right]
   \nonumber\\
    &+
    \frac{1}{z}\sum_{j=1}^{N_\mathrm{f}} \left[\mu_\mathrm{A}\bar{G}_{(j,\mathrm{A}),(j,\mathrm{A})}(z)-\mu_\mathrm{B}\bar{G}_{(j,\mathrm{B}),(j,\mathrm{B})}(z)\right].
    \label{eq:F_ssh_general}
\end{align}
A comparison between Eq.~(\ref{eq:F_ssh}) and Eq.~(\ref{eq:F_ssh_general}) indicates that 
$\bar{F}(z)$ has an additional term.
Also, $\bar{F}(z)$ is still related to the DOS $\rho_\mathrm{A}(E)$ and $\rho_\mathrm{B}(E)$:
\begin{align}
   &
   -\frac{1}{\pi}\mathrm{Im}\bar{F}(z=E+\mathrm{i}\eta)
   \nonumber\\
   =&
   \left(
      1+\frac{\mu_\mathrm{A}}{E}
   \right)
   \rho_\mathrm{A}(E)
   -
   \left(
      1+\frac{\mu_\mathrm{B}}{E}
   \right)
   \rho_\mathrm{B}(E).
   \label{eq:F_E_ssh_general}
\end{align}
From Eq.~\eqref{eq:F_E_ssh_general}, $\mathrm{Im}\bar{F}(E)$ diverges at $E=0$ as far as $\mu_\mathrm{A(B)}\neq0$ and $\rho_\mathrm{A(B)}(E)\propto {|E|}^{\alpha}$ with $0<\alpha<1$ for small $E$ [Fig.~\ref{fig:SSH}(b)].
This divergence does not 
originate
from  the non-zero value of the winding number.
It is notable that for $\mu_\mathrm{A}\neq0$ or $\mu_\mathrm{B}\neq0$, $\rho_\mathrm{A}(E)-\rho_\mathrm{B}(E)$ diverges but Eq.~\eqref{eq:F_E_ssh_general} converges according to the SBBC\@.
In Figs.~\ref{fig:SSH}(d) and (h), we show $\tilde{\rho}_\mathrm{A(B)}(E)$ corresponding to Fig.~\ref{fig:SSH}(b).
As explained, $\tilde{\rho}_\mathrm{A(B)}(E)$ does not have zero-energy peak [Figs.~\ref{fig:SSH}(d) and (h)] but $-\frac{1}{\pi}\mathrm{Im}F(E+\mathrm{i}\eta)$ has it [Fig.~\ref{fig:SSH}(b)].
In addition, in Fig.~\ref{fig:SSH}(h), there is an in-gap state at $E=-\mu_\mathrm{A}$ [see also Appendix~\ref{sec:App_SSH_ingap}].
This peak does not appear in Fig.~\ref{fig:SSH}(b) since the first term in Eq.~\eqref{eq:F_E_ssh_general} is zero at $E=-\mu_\mathrm{A}$.
Then the origins of the zero-energy peak in Figs.~\ref{fig:SSH}(a) and (b) are different: in Fig.~\ref{fig:SSH}(a), the origin is zero-energy peak in the DOS and in Fig.~\ref{fig:SSH}(b), it is the $\mu_\mathrm{A(B)}/E$ term in Eq.~\eqref{eq:F_E_ssh_general}.
In Figs.~\ref{fig:SSH}(f) and (j), $\rho_{\mathrm{A(B)}}^{\mathrm{bulk}}(E)$ is shown corresponding to (d) and (h), respectively and $\rho_{\mathrm{A}}^{\mathrm{bulk}}(E)\neq\rho_{\mathrm{B}}^{\mathrm{bulk}}(E)$ due to $\mu_\mathrm{A}\neq\mu_\mathrm{B}$.

If we choose
$\Gamma=\tau_1$ or $\tau_2$, $\bar{F}(z)$ contains off-diagonal elements of the Green's function, 
and it is difficult to obtain a concise physical interpretation.
Notably, the SSH model with chemical potential can be metallic, and the SBBC still holds.

\subsubsection{With impurity}
\begin{figure}[t]
   \centering
   \includegraphics[width=8.5cm]{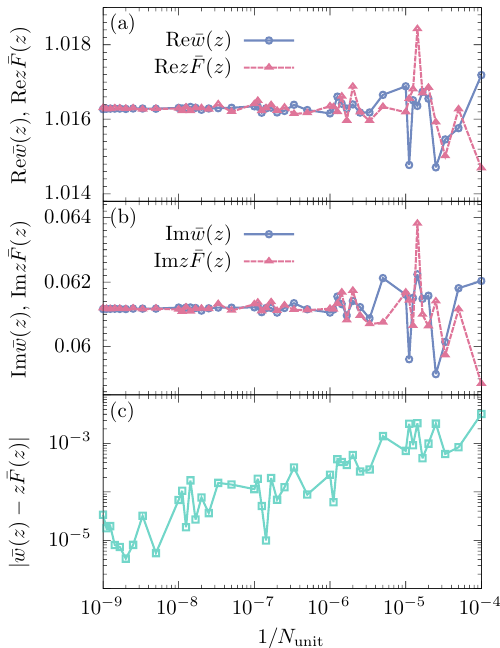}
   \caption{%
      (a) Real part and (b) imaginary part of $\bar{w}(z)$ and $z\bar{F}(z)$ plotted as a function of $1/N_\mathrm{unit}$
      for $t_2/t_1=1.5$, $\mu_\mathrm{A}/t_1=0.1$, $\mu_\mathrm{B}/t_1=0.2$, $z/t=0.1\imu$, $V_{\mu_\mathrm{A}}=0.5$, and $V_{\mu_\mathrm{B}}=0.8$.
      (c) $|\bar{w}(z)-z\bar{F}(z)|$ shown as a function of $1/N_\mathrm{unit}$
   }%
   \label{fig:SSH_imp}
\end{figure}

In Sec.~\ref{sec:kitaev_impurity}, the Kitaev chain with an impurity, which does not break the chiral symmetry, 
is considered. 
Here, we consider the SBBC for the SSH model with an onsite impurity potential, which breaks the chiral symmetry~\cite{PhysRevB.101.144204}.
The Hamiltonian is given by
\begin{align}
   {\cal H}_\mathrm{SSH}^{\mathrm{imp.}}
   =&{\cal H}_\mathrm{SSH}
   -\sum_j[\mu_{\mathrm{A}}+t_1f_{\mu_\mathrm{A}}(j)] n_{j,\mathrm{A}}\nonumber\\
    &-\sum_j[\mu_{\mathrm{B}}+t_1f_{\mu_\mathrm{B}}(j)] n_{j,\mathrm{B}}.
\end{align}
Here, $f_{\mu_{\mathrm{A(B)}}}$ is a random number that satisfies $|f_{\mu_{\mathrm{A(B)}}}|<V_{\mu_{\mathrm{A(B)}}}$.
In Fig.~\ref{fig:SSH_imp}, $\bar{w}(z)$ and $z\bar{F}(z)$ are shown
as a function of $N_\mathrm{unit}$.
$\bar{w}(z)$ and $z\bar{F}(z)$ converge to the same value,
and the difference between them becomes smaller as $N_\mathrm{unit}$
increases, as shown in the inset of Fig.~\ref{fig:SSH_imp}.
Although the proof given in Appendix~\ref{sec:App_proof_sbbc} has already provided firm evidence for the correspondence between the bulk and the edge, the system size dependence is not encoded in the proof.
Thus, numerical measurements are necessary, and they provide a quantitative estimate of the system size when the SBBC is applicable.
Parameter dependences of $\bar{w}(z)$ and $z\bar{F}(z)$ are also discussed in Appendix~\ref{sec:App_SSH_impurity}.

\section{Discussion\label{sec:discussion}}

Here, we discuss the possible applications of the SBBC\@.
As demonstrated in this study, the concept of SBBC can be widely used for a one-dimensional structure with correlation functions and is not restricted to chiral symmetric systems, a bulk with translational symmetry, or superconducting systems.
The essential aspects for applying the SBBC are to recognize the one-dimensional structure and divide the correlation function matrix into the commuting and anticommuting parts with a chiral operator (see Appendix~\ref{sec:App_proof_sbbc}).
For an illustration from a general perspective, we consider the single-particle function in equilibrium as follows: 
\begin{align}
G ( \sigma, \bm R_i, \tau_1 ; \sigma', \bm R_j, \tau_2 )
&= - \langle \mathcal T c_{i\sigma}(\tau_1) c^\dagger_{j\sigma'} (\tau_2)  \rangle,
\label{eq:Gdef}
\end{align}
where the index $i$ represents the three-dimensional spatial coordinate $\bm R_{i}$ of the lattice sites where the electrons are located.
We introduce the Heisenberg picture $A(\tau) = e^{\tau H} A e^{-\tau H}$ with imaginary time, and $\mathcal T$ is a time-ordering operator..
This correlation function should be considered for almost all the electronic systems of interest. For example, it can be calculated through first-principles calculations used to obtain electronic band structures.

Based on this example, we show that there can be, in principle, an infinite number of SBBC relations.
For simplicity, we assume that the bulk system has a translational symmetry.
First, we define the edge where the surface is perpendicular to the vector $\bm k_\perp$.
Then, we introduce vectors parallel to the surface $\bm k_{\parallel}$ that satisfy $\bm k_\perp \cdot \bm k_\parallel = 0$.
Similarly, in the spin space, we determine the chiral operator by choosing one of the directions $\bm \theta$, which is a unit vector specifying the direction of the spin, as
\begin{align}
\Gamma = \bm \theta \cdot \bm \sigma.
\end{align}
After the Fourier transformation, 
the bulk Green's function is written in a matrix form in the spin space as
\begin{align}
\bar g^{-1}(\bm k_\perp; \bm k_{\parallel}, \imu\omega_n ,\bm \theta)
= v + \alpha \Gamma + \bm \beta \cdot \bm \sigma,
\end{align}
where $\bm \theta \cdot \bm \beta = 0$.
From this expression, we can identify the commuting ($v+\alpha \Gamma$) and anticommuting ($\bm \beta \cdot \bm \sigma$) parts with the chiral operator.
The vector $\bm h = \alpha \bm \theta + \bm \beta$ can be regarded as a pseudomagnetic field in the $k$-space.
Note that the scalars $v,\alpha$ and the vector $\bm \beta$ are uniquely determined from the information of the system once $\bm \theta$ is specified.
Based on these expressions, we can express the bulk quantity as follows:
\begin{align}
\bar w &= 
\int
\frac{dk_\perp}{4\pi\imu}
\mathrm{Tr}\left(\Gamma \bar{g}\partial_\perp \bar{g}^{-1}\right)
\label{eq:defbulk}\\
       &=
\bm \theta \cdot
\int \frac{d k_\perp}{2\pi \imu} \,
\frac{
v \partial_\perp \bm h - \bm h \partial_\perp v - \imu \bm h \times \partial_\perp \bm h 
}{
v^2 - \bm h^2
}
\label{eq:genex} \\
&=
\int \frac{d k_\perp}{2\pi \imu} \,
\frac{
v \partial_\perp \alpha - \alpha \partial_\perp v - \imu \bm \theta \cdot (\bm \beta \times \partial_\perp \bm \beta)
}{
v^2 - \alpha^2 - \bm \beta^2
},
\end{align}
with $\partial_\perp = \frac{\partial}{\partial k_\perp}$.
Namely, the relations of the BBC are constructed for the {\it infinite} set of ($\bm k_{\parallel},\omega_n,\bm \theta$), although the counterpart quantity at the edge is not necessarily physically interpreted easily if the commuting part has a complicated structure, as shown in Eq.~\eqref{eq:def_F_j}.
The integrand of Eq.~\eqref{eq:defbulk} resembles an SU(2) spin-gauge field given by Eqs.~(155,156) of Ref.~\cite{TATARA2008213}, although Eq.~\eqref{eq:defbulk} is defined in the reciprocal space.
The appearance of the gauge-field-like structure is reminiscent of the relationship between the electromagnetic U(1) gauge field (i.e., scalar and vector potentials) and the Berry connection in the $\bm k$-space associated with the Bloch wave function~\cite{Vanderbilt_book}.

Thus, the appearance of the vector $\bm \theta$ in the spin space and the wave vector $\bm k_{\perp,\parallel}$ in the real (or reciprocal) space indicates that the SBBC is controlled by spin--orbit coupling.
Without complications in the spin and orbital spaces, the SBBC would provide a trivial relation of ``$0=0$'' between the bulk and the edge.
Hence, it is interesting to study a system with various types of spin--orbital couplings in detail, optionally combined with band-structure calculations.

For a system with a gap in the energy spectrum, the correlation length exhibits an exponential decay, which defines the spatially localized edge state well.
On the other hand, if we deal with the metallic state, the correlation functions generally decay with a power-law behavior, and the localized state at the edge is ill defined.
However, even in this case, the SBBC can be used if the spatial integral from the edge to the bulk converges.
For example, this point can be confirmed for the parameters at the topological critical point, where the correlation length diverges.

In the above discussion, we emphasize that we only invoke the presence of the edge and do not use any specific details of the system.
For the definition of the edge, the necessary condition is that the original Hamiltonian is composed of link variables with a finite range or decreasing amplitude when the distance increases.
This is a generic condition satisfied in most materials.
In addition, the extensions to the multiple degrees of freedom inside the unit cell, to the pairing states, and to two-particle correlation functions, such as charge and spin structure factors, are interesting and should be straightforward.
Furthermore, the expectation value in Eq.~\eqref{eq:Gdef} can be evaluated with many-body wave functions and are not restricted to one-body problems.
Thus, the SBBC should be widely applicable to general correlation functions, and it should be applied to extract nontrivial information at the edge from the bulk quantities.

\section{Summary and Outlook \label{sec:summary}}
In summary, we established
the generalization of the SBBC, which is an extended version of the bulk--boundary correspondence for topological systems.
We demonstrated the SBBC through several applications as follows.
(i) The SBBC holds for the Kitaev chain with impurities in the bulk, where the Hamiltonian has chiral symmetry.
(ii) In principle, the SBBC can be experimentally confirmed for the SSH model using the local DOS.
(iii) The SBBC holds for the SSH model with a chemical potential, which breaks the chiral symmetry.
(iv) The SBBC holds even if self-energy is included.

We suggest a possible application to a general correlation function in Sec.~\ref{sec:discussion}; however, several interesting problems remain. 
(1) The SBBC is only demonstrated for a one-dimensional system. Whether a similar demonstration is possible for a higher-dimensional system remains to be investigated.
(2) When $\sigma_{\mathrm{c}}(z)=0$ in Eq.~(\ref{eq:bar_w}), $\lim_{z\rightarrow0}\bar{w}(z)$ is an integer. The relationship between this integer and the number of edge modes for many-body interaction systems remains unexplored.
(3) In this study, we adopted Eq.~(\ref{eq:w_real_space}) and Eq.~(\ref{eq:bar_w}) as an extended version of the winding number in a real-space representation.
Recently, another representation of the winding number related to the Bott index has been suggested~\cite{PhysRevB.103.224208}.
Its possible extension to a nonzero frequency is an interesting problem to be investigated.
(4) 
We only consider $\bar{G}(z)$ in Eq.~(\ref{eq:g_general}) as the Green's function. On the other hand, it can be regarded as a general matrix, such as a spin--spin correlation function. This is also an interesting issue.

Thus, the present study presented a general framework for bulk--boundary correspondence.
Numerous target systems in which the concept of the SBBC is applicable exist, and they remain to be explored using mathematical and computational techniques.

\begin{acknowledgments}
   We are grateful to A.~Daido, S.~Richard, C.~Bourne, M.~Lein and H.~Katsura
   for useful discussions.
   This work was supported by 
   Grant-in-Aid for Scientific Research B (KAKENHI Grant No.~JP18H01176 and No.~JP20H01857) 
   from the Ministry of Education, Culture, Sports, Science, and Technology, Japan (MEXT)
   and  Grant-in-Aid for Scientific Research A (KAKENHI Grant No.~JP20H00131), the JSPS Core-to-Core program Oxide Superspin International Network and Researcher Exchange Program between JSPS and RFBR (JPJSBP120194816).
   This work was also supported by Japan Society for Promotion of Science (JSPS) KAKENHI Grant No.~18K13490.
\end{acknowledgments}

\bibliography{biblio}

\begin{thebibliography}{48}
\bibitem [{chi()}]{chiral}%
  \BibitemOpen
  \href@noop {} {}\bibinfo {note} {$\mathrm{Tr}\Gamma=0$ is always satisfied
  when the eigenvalues of the Hamiltonian does not include zero:
  $\mathrm{Tr}\Gamma=\mathrm{Tr}\Gamma HH^{-1}=-\mathrm{Tr}H\Gamma
  H^{-1}=-\mathrm{Tr}\Gamma$.}\BibitemShut {Stop}%
\bibitem [{Gam()}]{Gamma}%
  \BibitemOpen
  \href@noop {} {}\bibinfo {note} {$\Gamma^\dagger=\Gamma$ and $\Gamma^2=1$ are
  necessary to obtain integer $W$. On the other hand, Eq.~(\ref{eq:sbbc}) holds
  without these conditions.}\BibitemShut {Stop}%
\bibitem [{not()}]{note_for_x}%
  \BibitemOpen
  \href@noop {} {}\bibinfo {note} {In Eq.~(\ref{eq:w_real_space}), there
  appears $r_i-r_j$ from $[g(z),\bar{x}]$ and it is noted that
  $r_1-r_{N_\mathrm{unit}}=1$ due to the periodic boundary condition. Then the
  winding number $w(z=0)$ becomes $w(0)=\frac{1}{2\imu
  N_\mathrm{unit}}\bar{\mathrm{Tr}}\{\bar{\Gamma}h^{-1}[h,\bar{x}]\} \neq
  \frac{1}{\imu N_\mathrm{unit}}\bar{\mathrm{Tr}}(\bar{\Gamma}\bar{x})$. It is
  also noted that for superconductors, $\bar{x}$ does not have physical
  meaning.}\BibitemShut {Stop}%
\end{thebibliography}
\appendix

\section{Derivation of Eq.~\eqref{eq:w_surface}\label{sec:App_der_w}}
   Here we derive Eq.~\eqref{eq:w_surface} from Eq.~\eqref{eq:w_real_space}.
   We consider the correspondence between $N_\mathrm{unit}$ unit cells system with the open and periodic boundary condition.
   For the system with open boundary condition, we assume the edges are located at $j=1$ and $j=N_\mathrm{unit}$.
   We set $t=h_{j,j+1}=H_{j,j+1}$ and assume translational symmetry (no impurity potential).
   Without loss of generality, we set $h_{j,j+l}=H_{j,j+l}=0$ for $l\geq2$.
   Eq.~\eqref{eq:w_real_space} is
   \begin{align}
      w(z)
      =&
      -
      \frac{1}{2\mathrm{i}
      N_\mathrm{unit}}
      \bar{\mathrm{Tr}}
      \left\{
         \bar{\Gamma}
         g(z)\mathrm{i}[h,\bar{x}]
      \right\}.
   \end{align}
   By using translational symmetry, we obtain
   \begin{align}
      &
      -\frac{1}{2\mathrm{i}N_\mathrm{unit}}
      \bar{\mathrm{Tr}}
      \left\{
         \bar{\Gamma}
         g(z)\mathrm{i}[h,\bar{x}]
      \right\}
      \nonumber\\
      =&
      -\frac{1}{2\mathrm{i}}
      \mathrm{Tr}
      \left(
         \Gamma
         \left\{
            g_{1,2}(z)\mathrm{i}[h,\bar{x}]_{2,1}
            +
            g_{2,1}(z)\mathrm{i}[h,\bar{x}]_{1,2}
         \right\}
      \right)
      \nonumber\\
      =&
      \frac{1}{2}
      \mathrm{Tr}
      \left\{
         \Gamma
         \left[
            g_{1,2}(z)t^\dagger
            -
            g_{2,1}(z)t
         \right]
      \right\}.
   \end{align}
   Utilizing the following relations that hold for $N_\mathrm{unit}\rightarrow\infty$\cite{PhysRevB.101.024509}
   \begin{align}
      g_{1,2}(z)
      =&
      g_{1,1}(z)tG_{1}
      =
      G_{N} t g_{1,1}(z),
      \label{eq:g12_app}
      \\
      g_{2,1}(z)
      =&
      g_{1,1}(z)t^\dagger G_{N}
      =
      G_{1}t^\dagger g_{1,1}(z),
      \label{eq:g21_app}
      \\
      g_{1,1}(z)
      =&
      {(1-G_N t G_1 t^\dagger)}^{-1}G_N
      \nonumber
      \\
      =&
      {(1-G_1 t^\dagger G_N t)}^{-1}G_1,
      \label{eq:g11_app}
   \end{align}
   with $G_1=G_{1,1}(z)$ and $G_N=G_{N_\mathrm{unit}N_\mathrm{unit}}(z)$,
   $w(z)$ can be written as
   \begin{align}
      w(z)=&
      \frac{1}{2}
      \mathrm{Tr}
      \left[
         \Gamma
         g_{1,1}(z)
         \left(
            tG_1 t^\dagger
            -
            t^\dagger G_N t
         \right)
      \right]
      \nonumber\\
      =&
      \frac{1}{2}
      \mathrm{Tr}
      \left[
         \Gamma
         g_{1,1}(z)
         \left(
            -G_N^{-1}
            +tG_1 t^\dagger
            +
            G_N^{-1}
            \right.
            \right.
            \nonumber\\
       &
       \left.
       \left.
          \hspace{2.1cm}
            +
            G_1^{-1}
            -
            t^\dagger G_N t
            -
            G_1^{-1}
         \right)
      \right]
      \nonumber\\
      =&
      \frac{1}{2}
      \mathrm{Tr}
      \left[
         \Gamma
         g_{1,1}(z)
         \left(
            G_N^{-1}
            -
            G_1^{-1}
         \right)
      \right]
      \nonumber\\
      =&
      \frac{1}{2}
      \mathrm{Tr}
      \left\{
         \Gamma
         \left[
            {(1-G_N tG_1t^\dagger)}^{-1}
            -
            {(1-G_1t^\dagger G_N t)}^{-1}
         \right]
      \right\}.
   \end{align}
   Here, Eqs.~\eqref{eq:g12_app}--\eqref{eq:g11_app} connect the system with periodic boundary condition to the system with open boundary condition.
   From the second to the third line, we use Eq.~\eqref{eq:g11_app}.
   We note that, in the above calculations, we do not use the anticommutation relation between the chiral operator and the Hamiltonian.

\section{Proof of generalized SBBC\label{sec:App_proof_sbbc}}

In this appendix,
most of the quantities denoted by upper-case (lower-case) letters are defined in the open (periodic) boundary condition system and we use a bit different notation from the main text.

We follow the same procedure as in the Daido and Yanase's paper~\cite{PhysRevB.100.174512} to
prove Eq.~(\ref{eq:sbbc_general}):
\begin{align}
   \bar{\cal F}=&\bar{\mathfrak w}
\label{eq:super_general}
\end{align}
with
\begin{align}
   \bar{\cal F}=&
   \frac{1}{N_\mathrm{diff}}
   \sum_{j=N_\mathrm{i}}^{N_\mathrm{f}}
   {\cal F}_j,
   \\
   {\cal F}_j=&
   \sum_{l=1}^j
   \bar{\mathrm{Tr}}{\left[{\cal N}_l D_\mathrm{c}\bar{\Gamma} \bar{\cal G}\right]},
   \label{eq:F_j_general}
   \\
   \bar{\mathfrak w}=&
   \frac{1}{2\imu N_\mathrm{unit}}
   \bar{\mathrm{Tr}}\{\bar{\Gamma} \bar{\mathfrak g}\imu[\bar{\mathfrak g}^{-1},\bar{x}]\},
\end{align}
with $N_{\mathrm{diff}}=N_\mathrm{i}-N_\mathrm{f}+1$ and $\bar{x}$ given by Eq.~(\ref{eq:position}).
Here $\bar{\cal F}$ is defined in the system with the open boundary condition at the both edges,
$\bar{\mathfrak w}$ is defined in the system with the periodic boundary condition.
Here, $\bar{\cal G}$ and $\bar{\mathfrak g}$ are given by
\begin{align}
   \bar{\cal G}
   =&
   {[D_\mathrm{c}+D_\mathrm{ac}]}^{-1},
   \label{eq:app_commute}
   \\
   \bar{\mathfrak g}
   =&
   {[d_\mathrm{c}+d_\mathrm{ac}]}^{-1},
   \label{eq:app_anticommute}
\end{align}
where $D_{\mathrm{c,ac}}$ and $d_{\mathrm{c,ac}}$ satisfy following relations:
\begin{align}
   0=&[\bar{\Gamma}, D_\mathrm{c}]
   =\{\bar{\Gamma}, D_\mathrm{ac}\},
   \\
   0=&[\bar{\Gamma}, d_\mathrm{c}]
   =\{\bar{\Gamma}, d_\mathrm{ac}\}.
\end{align}
Here we assume $D_\mathrm{c}+D_\mathrm{ac}$, $D_\mathrm{c}$, 
$d_\mathrm{c}+d_\mathrm{ac}$ and $d_\mathrm{c}$ are regular matrices and its matrix elements are short range in a real space basis.
For example, Eq.~(\ref{eq:g_general}) with $z\bar{F}(z)=\bar{\cal F}$ and $\bar{w}(z)=\bar{\mathfrak w}$ is reproduced by setting
\begin{align}
   D_\mathrm{c} =& z-\Sigma_\mathrm{c}(z),
   \\
   D_\mathrm{ac} =&-H -\Sigma_\mathrm{ac}(z).
\end{align}
Then Eq.~(\ref{eq:F_j_general}) can be written as
\begin{align}
   {\cal F}_j
   =&
   \sum_{l=1}^j
   \bar{\mathrm{Tr}}{\left\{{\cal N}_l D_\mathrm{c}\bar{\Gamma} {\left[D_\mathrm{c}+D_\mathrm{ac}\right]}^{-1}\right\}}
   \nonumber\\
   =&
   \sum_{l=1}^j
   \bar{\mathrm{Tr}}{\left\{{\cal N}_l\bar{\Gamma} {\left[1+D_\mathrm{ac}D_\mathrm{c}^{-1}\right]}^{-1}\right\}}.
   \label{eq:1_CC}
\end{align}
${[1+D_\mathrm{ac}D_\mathrm{c}^{-1}]}^{-1}$ in Eq.~(\ref{eq:1_CC}) is similar to ${(z-H)}^{-1}$
where $H$ is replaced by $D_\mathrm{ac}D_\mathrm{c}^{-1}$.
Recognizing this correspondence and then following the Daido and Yanase's paper~\cite{PhysRevB.100.174512}, we can prove the generalized SBBC\@.

\section{$V_\mu$ and $\mu$ dependence of $\bar{w}(z)$ and $z\bar{F}(z)$ for Kitaev chain with impurity potential\label{sec:App_Kitaev}}
\begin{figure}[H]
   \centering
   \includegraphics[width=8.5cm]{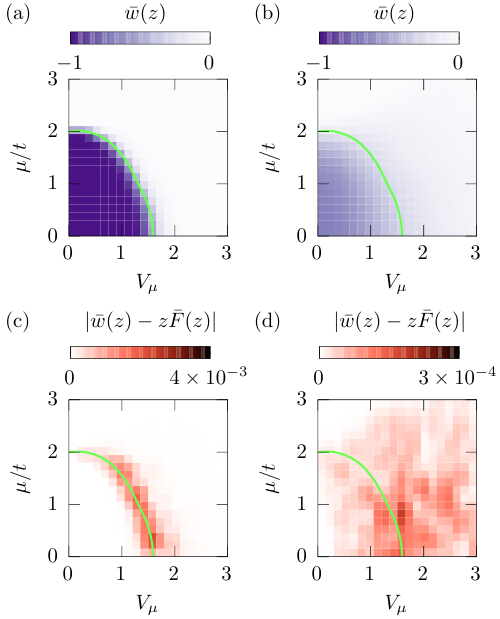}
   \caption{%
      Real part of $\bar{w}(z)$ [(a) and (b)] and $|\bar{w}(z)-z\bar{F}(z)|$ [(c) and (d)] are plotted as functions of $\mu$ and $V_\mu$ for $\Delta/t=0.1$ and $N_\mathrm{unit}=10^8$. 
      $z/t=10^{-4}\mathrm{i}$ for (a) and (c) and $z/t=10^{-1}\mathrm{i}$ for (b) and (d).
      Green curve indicates the phase boundary.
      Imaginary part of $\bar{w}(z)$ with purely imaginary $z$ is zero.
   }%
   \label{fig:phase_diagram_Kitaev}
\end{figure}
We show $\bar{w}(z)$ for the Kitaev chain in Figs.~\ref{fig:phase_diagram_Kitaev}(a) and (b) for two values of $z$ [$z/t=10^{-4}\mathrm{i}$ for Fig.~\ref{fig:phase_diagram_Kitaev}(a) and $10^{-1}\mathrm{i}$ for (b)].
A green curve is the phase boundary between the topologically nontrivial and the gapless phase determined by a transfer matrix method~\cite{PhysRevB.93.075129} [see also Fig.~\ref{fig:phase_kitaev}].
For the small value of $z$ [Fig.~\ref{fig:phase_diagram_Kitaev}(a)], $\bar{w}(z)$ becomes almost $-1$ in the topologically nontrivial phase since $\bar{w}(z)$ becomes the winding number $W$ for $z\rightarrow0$ in this phase.
On the other hand, $\bar{w}(z)$ is almost zero in other phases.
For a larger value of $z$ [Fig.~\ref{fig:phase_diagram_Kitaev}(b)], $\bar{w}(z)$ in the topologically nontrivial phase is larger than $-1$ but the value is still smaller than that in the other phases.

The difference between $\bar{w}(z)$ and $z\bar{F}(z)$ is shown in Figs.~\ref{fig:phase_diagram_Kitaev}(c) and (d).
This difference is basically small in the entire region but is a bit larger close to the topological transition point for small value of $z$ [Fig.~\ref{fig:phase_diagram_Kitaev}(c)].
The difference becomes smaller when the larger system size is taken.

\section{Energy gap and DOS for Kitaev chain with impurity potential\label{sec:App_Kitaev_dos}}
\begin{figure}[H]
   \centering
   \includegraphics[width=8.5cm]{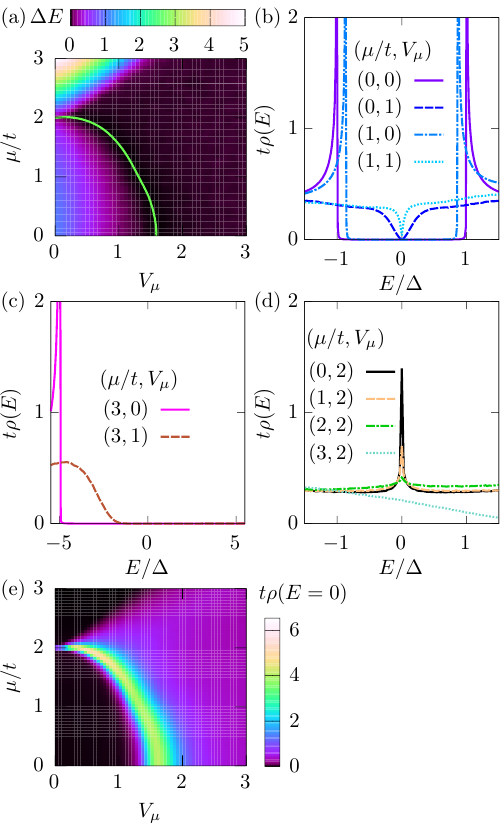}
   \caption{%
      (a) Energy gap at Fermi energy normalized by $2\Delta$ is plotted as functions of $V_\mu$ and $\mu$
      estimated from the diagonalization of $N_{\mathrm{unit}}=2000$ with 10 sample average is shown.
      Green curve is the phase boundary estimated from transfer matrix method~\cite{PhysRevB.93.075129}.
      DOS $\rho(E)$ with $\eta/t=10^{-4}$ for (b) topologically nontrivial phase, (c) topologically trivial phase and (d) gapless phase are plotted as a function of $E$ with $N_\mathrm{unit}=10^8$ calculated by using recursive Green's function method~\cite{PhysRevB.55.5266} with \cite{eigen1977}.
         (e) DOS at zero energy $\rho(E=0)$ with $\eta/t=10^{-4}$ is plotted as functions of $V_\mu$ and $\mu$ with $N_\mathrm{unit}=10^7$.
      In all graphs, we set $\Delta/t=0.1$.
   }%
   \label{fig:app_dos_kitaev_imp}
\end{figure}
In relation to Fig.~\ref{fig:phase_kitaev}, we show the actual energy gap at Fermi energy in Fig.~\ref{fig:app_dos_kitaev_imp}(a) and the DOS for particle part 
\begin{align}
   \rho(E)=-\frac{1}{\pi N_\mathrm{unit}}\mathrm{Im}\bar{\mathrm{Tr}}\left[P\bar{g}(E+\mathrm{i}\eta)\right],
\end{align}
with $P=\frac{1}{2}\mathrm{diag}(\tau_0+\tau_3,\tau_0+\tau_3,\ldots)$
in Fig.~\ref{fig:app_dos_kitaev_imp}(b), (c) and (d) for the periodic system.
   It is noted that the energy gap in the topologically nontrivial phase is determined mainly from the superconducting energy gap $\Delta$ [see also Fig.~\ref{fig:app_dos_kitaev_imp}(b)].
On the other hand, in the topologically trivial phase, the energy gap is related to the band top or bottom [see also Fig.~\ref{fig:app_dos_kitaev_imp}(c)].
Then, in the topologically nontrivial phase, the energy gap is smaller or equal to $\Delta$ and in the topologically trivial phase, it is mainly determined by $\mu$.
\\ \indent
In the gapless phase [Fig.~\ref{fig:app_dos_kitaev_imp}(d)], we can confirm there is no energy gap and
there is zero energy peak for $\mu/t=0$, $1$, and $2$.
This zero energy peak is observed close to the phase boundary between the topologically nontrivial phase and the gapless phase [Fig.~\ref{fig:app_dos_kitaev_imp}(e)].
The low-energy peak for $V_\mu>0$ might originate from the Majorana fermions in the bulk of the disordered media.
Namely, in the presence of the impurity potentials, the one-dimensional chain is effectively separated into segments, and each segment has Majorana edge modes at its edges depending on the parameters.
Since these Majorana states show the low-energy peak in DOS, their accumulation results in Fig.~\ref{fig:app_dos_kitaev_imp}(d).

\section{Inverse participation ratio for Kitaev chain with impurity potential\label{sec:App_Kitaev_IPR}}
\begin{figure}[H]
   \centering
   \includegraphics[width=8.5cm]{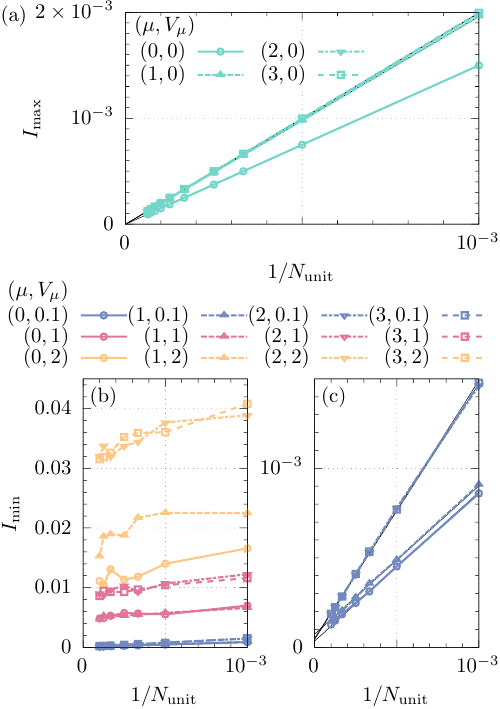}
   \caption{%
      (a)$I_\mathrm{max}$ is plotted as a function of $N_\mathrm{unit}$ for $V_\mu=0$ with several values of $\mu$.
      (b)$I_\mathrm{min}$ with 10 samples average is plotted as a function of $N_\mathrm{unit}$ for $V_\mu>0$ with several values of $\mu$.
      (c)magnified view of (b) is shown.
      In (a) and (c), black lines are eye guides.
   }%
   \label{fig:kitaev_ipr}
\end{figure}

We discuss an inverse participation ratio (IPR) in this appendix which characterizes localization length of the wave function.
The IPR is
\begin{align}
   I_\nu=&
   \frac{\sum_{j=1}^{N_\mathrm{unit}}{|\psi_\nu(j)|}^4}{{\left[\sum_{j=1}^{N_\mathrm{unit}}{|\psi_\nu(j)|}^2\right]}^2},
\end{align}
where $\psi_\nu(j)$ is the eigen vector of the eigen value $E_\nu$ of the Hamiltonian with the periodic boundary condition.
If $I_\nu\rightarrow0$ for $N_\mathrm{unit}\rightarrow\infty$, $\psi_\nu(j)$ is delocalized and if $I_\nu$ approaches non-zero value, $\psi_\nu(j)$ is localized.
In Fig.~\ref{fig:kitaev_ipr}(a), we show $I_{\max}=\max(I_\nu)$ for $V_\mu=0$ for the Kitaev chain with impurity potential.
We can see that $I_{\max}$ is zero for $N_\mathrm{unit}\rightarrow\infty$.
Then all the wave functions are not localized for $V_\mu=0$.
In Figs.~\ref{fig:kitaev_ipr}(b) and (c), we show $I_{\min}=\min(I_\nu)$ for $V_\mu>0$.
In all plots, $I_{\min}$ is not zero for $N_\mathrm{unit}\rightarrow\infty$.
Then we can conclude that in all phases with $V_\mu>0$, all the wave functions are localized.

\section{$\rho_\mathrm{A}(E)-\rho_\mathrm{B}(E)$ for SSH model without chemical potential\label{sec:App_SSH}}
Analytical expression of the difference of the DOS for the SSH model with $\mu_\mathrm{A}=\mu_\mathrm{B}=0$ is given by
\begin{align}
   &
   \rho_\mathrm{A}(E)-\rho_\mathrm{B}(E)
   =
   -
   \frac{1}{\pi}\mathrm{Im}
   [w(E+\imu\eta)/(E+\imu\eta)],
   \\
   &w(E+\imu\eta)
   \nonumber\\
   =&
   \frac{1}{2}
   \left[
      1
      -\frac{[t_1^2-t_2^2-{(E-\imu\eta)}^2]{(-1)}^{{\mathfrak f}(t_1,t_2,E+\imu\eta)}}
      {\sqrt{{[t_1^2+t_2^2-{(E+\imu\eta)}^2]}^2-{(2t_1t_2)}^2}}
   \right],
\end{align}
with
\begin{align}
   &{\mathfrak f}(t_1,t_2,z)
   =
   \max\{n\in\mathbb{Z}\mid n\leq \tilde{\mathfrak f}(t_1,t_2,z)\},
   \\
   &\tilde{\mathfrak f}(t_1,t_2,z)
   =\frac{1}{2\pi}\left(\pi-\arg[t_1-t_2-z]+\arg[t_1+t_2-z]\right.
   \nonumber \\
    &\left.\hspace{19mm}-\arg[t_1-t_2+z]+\arg[t_1+t_2+z]\right).
\end{align}

\section{Local Green's function and in-gap state of SSH model with chemical potential\label{sec:App_SSH_ingap}}
Here, we discuss the chemical potential dependence of an in-gap state of SSH model without impurity potential.
The local Green's function for right semi-infinite system is~\cite{PhysRevB.99.184512,PhysRevB.101.024509}
\begin{align}
   G_{(j,\mathrm{A}),(j,\mathrm{A})}(z)
   =&
   {[\tilde{G}^{-1}_{j-1}(z)-t_\mathrm{1}^2G_\infty(z)]}^{-1},
   \label{eq:GAA_SSH}
   \\
   G_{(j,\mathrm{B}),(j,\mathrm{B})}(z)
   =&
   {[G_\infty^{-1}(z)-t_\mathrm{1}^2\tilde{G}_{j-1}(z)]}^{-1},
   \label{eq:GBB_SSH}
\end{align}
with
\begin{widetext}
   \begin{align}
      \tilde{G}_j(z)
      =&
      \frac{1}{z+\mu_\mathrm{A}}
      \frac{(\alpha_{--}^j\alpha_{++}-\alpha_{-+}^j\alpha_{+-})+2(z+\mu_\mathrm{A})(z+\mu_\mathrm{B})(\alpha_{-+}^j-\alpha_{--}^j)}
      {2t_1^2(\alpha_{--}^j-\alpha_{-+}^j)+(\alpha_{-+}^j\alpha_{++}-\alpha_{--}^j\alpha_{+-})},
      \\
      G_\infty(z)
      =&
      \lim_{n\rightarrow\infty}
      2\frac{(z+\mu_\mathrm{A})(-\alpha_{--}^n+\alpha_{-+}^n)}{-\alpha_{--}^n\alpha_{+-}+\alpha_{-+}^n\alpha_{++}}
      =
      \begin{cases}
         2(z+\mu_\mathrm{A})/\alpha_{+-} & |\alpha_{--}|>|\alpha_{-+}|,
         \\
         2(z+\mu_\mathrm{A})/\alpha_{++} & |\alpha_{--}|<|\alpha_{-+}|,
      \end{cases}
      \\
      \alpha_{\pm\pm}
      =&
      \pm t_\mathrm{1}^2 - t_\mathrm{2}^2 + (z+\mu_\mathrm{A})(z+\mu_\mathrm{B})
      \pm \sqrt{-4t_\mathrm{1}^2 t_\mathrm{2}^2 + {[t_\mathrm{1}^2+t_\mathrm{2}^2-(z+\mu_\mathrm{A})(z+\mu_\mathrm{B})]}^2}.
   \end{align}
\end{widetext}
Here $j=1$ indicates the left surface unit cell.
From Eq.~\eqref{eq:GAA_SSH}, we can check that $G_{(j,\mathrm{A}),(j,\mathrm{A})}(z)$ diverges at $z=-\mu_\mathrm{A}$ when $|t_\mathrm{1}/t_\mathrm{2}|<1$.
Hence, there is the in-gap state at $E=-\mu_\mathrm{A}$ when $|t_\mathrm{1}/t_\mathrm{2}|<1$.
Note that there is no in-gap state at B sublattice from Eq.~\eqref{eq:GBB_SSH}.

\section{Parameters dependence of $\bar{w}(z)$ and $z\bar{F}(z)$ for SSH model with chemical potential\label{sec:App_SSH_impurity}}
\begin{figure}[H]
   \centering
   \includegraphics[width=8.5cm]{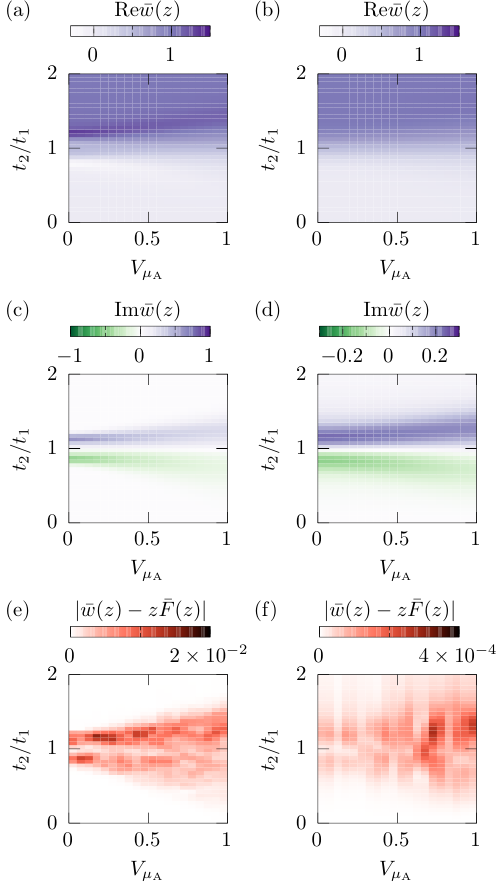}
   \caption{%
      $\bar{w}(z)$ and $|\bar{w}(z)-z\bar{F}(z)|$ are plotted as functions of $t_2/t_1$ and $V_{\mu_\mathrm{A}}$.
      In (a) and (b), real part of $\bar{w}(z)$ is shown. In (c) and (d), imaginary part of $\bar{w}(z)$ is shown.
      In (e) and (f), $|\bar{w}(z)-z\bar{F}(z)|$ is shown.
      $z/t_1=10^{-4}\mathrm{i}$ for (a), (c) and (e).
      $z/t_1=10^{-1}\mathrm{i}$ for (b), (d) and (f).
      $\mu_\mathrm{A}/t_1=0.1$, $\mu_\mathrm{B}/t_1=0.2$, $V_{\mu_\mathrm{B}}=0.8$, and $N_\mathrm{unit}=10^8$
   }%
   \label{fig:app_ssh_impuriry}
\end{figure}

We show $t_2$ and $V_{\mu_\mathrm{A}}$ dependence of $\bar{w}(z)$ and $z\bar{F}(z)$ for the SSH model with impurity potential in Fig.~\ref{fig:app_ssh_impuriry}.
The data supports the SBBC in the disordered SSH model.

In order to gain more insights,
let us consider two extreme limits: $t_1=0$ or $t_2=0$.
In this appendix, we define $\mu_{l,\mathrm{A(B)}}=\mu_{\mathrm{A(B)}}+t_1f_{\mathrm{A(B)}}(l)$ for a short-hand notation.
First, when $t_1=0$, the bond between the intra unit cell vanishes and the surface sites are completely isolated for the open boundary system.
The local Green's function for $l=1$ is
\begin{align}
   G_{1,1}(z)
   =&
   \begin{pmatrix}
      \frac{1}{z+\mu_{1,\mathrm{A}}}&0
      \\
      0&
      \frac{z+\mu_{2,\mathrm{A}}}{(z+\mu_{2,\mathrm{A}})(z+\mu_{1,\mathrm{B}})-t_{2}^{2}}
   \end{pmatrix}.
\end{align}
Also, for $1<l<N_\mathrm{unit}$, the local Green's function is
\begin{align}
   G_{l,l}(z)
   =&
   \begin{pmatrix}
      \frac{z+\mu_{l-1,\mathrm{B}}}{(z+\mu_{l,\mathrm{A}})(z+\mu_{l-1,\mathrm{B}})-t_{2}^{2}} & 0
      \\
      0 & \frac{z+\mu_{l+1,\mathrm{A}}}{(z+\mu_{l+1,\mathrm{A}})(z+\mu_{l,\mathrm{B}})-t_{2}^{2}}
   \end{pmatrix}.
\end{align}
Then, $zF_j(z)$ with $j>1$ is given by
\begin{align}
   zF_j(z)
   =&
   \sum_{l=1}^j
   \mathrm{Tr}\tau_3 M_lG_{l,l}(z)
   \nonumber\\
   =&
   1
   -\frac{(z+\mu_{j+1,\mathrm{A}})(z+\mu_{j,\mathrm{B}})}{(z+\mu_{j+1,\mathrm{A}})(z+\mu_{j,\mathrm{B}})-t_2^2},
   \label{eq:F_ssh_t1_0}
\end{align}
with $M_l=\mathrm{diag}(z+\mu_{l,\mathrm{A}},z+\mu_{l,\mathrm{B}})$.
Here, it is notable that $zF_j(z)$ only depends on $\mu_{j,\mathrm{B}}$ and $\mu_{j+1,\mathrm{A}}$.
Then $\lim_{z\rightarrow0}z\bar{F}(z)=\lim_{z\rightarrow0}z\langle F_j(z)\rangle_j$ deviates from unity when the chemical potentials are not zero.
If $\mu_{j+1,\mathrm{A}}\mu_{j,\mathrm{B}}/t_2^2\ll1$ is satisfied, $\lim_{z\rightarrow0}z\bar{F}(z)$ is approximately unity.
It is also notable that when $\mu_{j,\mathrm{A}}\neq0$ or $\mu_{j,\mathrm{B}}\neq0$ are satisfied only for $j<N_\mathrm{i}$, $z\bar{F}(z)$ is unity due to Eq.~\eqref{eq:F_ssh_t1_0}.
In this case, although there is no zero-energy state protected by chiral symmetry since chiral symmetry is broken, $\lim_{z\rightarrow0}z\bar{F}(z)$ is the nonzero integer.

Next, when $t_2=0$, the bond between the unit cells vanishes and open and periodic boundary conditions can not be distinguished.
Also, at $t_2=0$, we can see $\bar{w}(z)=z\bar{F}(z)=0$ independent of the values of chemical potentials [see also Fig~\ref{fig:app_ssh_impuriry}(a)--(d)].
Due to $[\bar{h},\bar{x}]=0$ where $\bar{h}$ is the Hamiltonian for the SSH model with the chemical potentials $\mu_\mathrm{A}$, $\mu_\mathrm{B}$ and impurity potential, we obtain $\bar{w}(z)=0$.
The local Green's function is
\begin{align}
   G_{l,l}(z)
   =&
   {%
      \begin{pmatrix}
         z+\mu_{l,\mathrm{A}}&t_{1}
         \\
         t_1&z+\mu_{l,\mathrm{B}}
      \end{pmatrix}
   }^{-1}
   \nonumber\\
   =&
   \frac{1}{(z+\mu_{l,\mathrm{A}})(z+\mu_{l,\mathrm{B}})-t_{1}^{2}}
   \begin{pmatrix}
      z+\mu_{l,\mathrm{B}} & -t_{1}
      \\
      -t_1 & z+\mu_{l,\mathrm{A}}
   \end{pmatrix}.
\end{align}
Then $zF_j(z)$ is
\begin{align}
   zF_j(z)
   =&
   \sum_{l=1}^j
   \mathrm{Tr}
   \tau_3
   M_lG_{l,l}(z)
   =0.
\end{align}
Thus, we obtain $z\bar{F}(z)=z\langle F_j(z)\rangle_j=0$.

\end{document}